\documentclass{IEEEoj}

\usepackage{tikz}
\newcommand{\circnum}[1]{%
  \tikz[baseline=(char.base)]%
    \node[draw,circle,inner sep=0.5pt,font=\small](char){#1};%
}

\usepackage{amsmath,amsfonts}
\usepackage{algorithmicx}
\usepackage{algpseudocode}
\usepackage{algorithm}
\usepackage{array}
\usepackage[caption=false,font=normalsize,labelfont=sf,textfont=sf]{subfig}
\usepackage{textcomp}
\usepackage[utf8]{inputenc}
\usepackage{stfloats}
\usepackage{url}
\usepackage{verbatim}
\usepackage{booktabs}
\usepackage{graphicx}
\usepackage{cite}
\usepackage{float}
\usepackage{multirow}
\usepackage{listings}
\usepackage[most]{tcolorbox}
\usepackage{caption}
\usepackage{colortbl}
\usepackage{amssymb}
\usepackage{rotating}
\usepackage{threeparttable}
\newcommand{\xmark}{--}
\usepackage{pifont}  
\usepackage{setspace}
\newcommand{\cmark}{\checkmark} 
\DeclareCaptionFormat{listing}{\vspace{-0.1cm}#1#2#3}
\def\BibTeX{{\rm B\kern-.05em{\sc i\kern-.025em b}\kern-.08em
    T\kern-.1667em\lower.7ex\hbox{E}\kern-.125emX}}
\AtBeginDocument{\definecolor{ojcolor}{cmyk}{0.93,0.59,0.15,0.02}}
\def\OJlogo{\vspace{-4pt}\hskip-4pt\includegraphics[height=18pt]{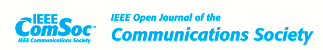}}

\begin{document}
\receiveddate{XX Month, XXXX}
\reviseddate{XX Month, XXXX}
\accepteddate{XX Month, XXXX}
\publisheddate{XX Month, XXXX}
\currentdate{11 January, 2024}
\doiinfo{OJCOMS.2024.011100}


\title{NetIntent: Leveraging Large Language Models for End-to-End Intent-Based SDN Automation}

\author{Md. Kamrul Hossain\IEEEauthorrefmark{1}, Walid Aljoby\IEEEauthorrefmark{1}}
\affil{Information and Computer Science Department, King Fahd University of Petroleum \& Minerals (KFUPM), Dhahran, Saudi Arabia}
\corresp{CORRESPONDING AUTHOR: Walid Aljoby (e-mail: waleed.gobi@kfupm.edu.sa).}
\markboth{Preparation of Papers for IEEE OPEN JOURNALS}{Author \textit{et al.}}



\begin{abstract}
Intent-Based Networking (IBN) often leverages the programmability of Software‑Defined Networking (SDN) to simplify network management. However, significant challenges remain in automating the entire pipeline, from user-specified high-level intents to device-specific low-level configurations.
Existing solutions often rely on rigid, rule-based translators and fixed APIs, limiting extensibility and adaptability. By contrast, recent advances in large language models (LLMs) offer a promising pathway that leverages natural language understanding and flexible reasoning.
However, it is unclear to what extent LLMs can perform IBN tasks.
To address this, we introduce \textit{\textbf{IBNBench}}, a first-of-its-kind benchmarking suite comprising four novel datasets: Intent2Flow-ODL, Intent2Flow-ONOS, FlowConflict-ODL, and FlowConflict-ONOS. 
These datasets are specifically designed for evaluating LLMs performance in intent translation and conflict detection tasks within the industry-grade SDN controllers ODL and ONOS.
Our results provide the first comprehensive comparison of 33 open-source LLMs on IBNBench and related datasets, revealing a wide range of performance outcomes.
However, while these results demonstrate the potential of LLMs for isolated IBN tasks, integrating LLMs into a fully autonomous IBN pipeline remains unexplored. Thus, our second contribution is \textbf{\textit{NetIntent}}, a unified and adaptable framework that leverages LLMs to automate the full IBN lifecycle, including translation, activation, and assurance within SDN systems. NetIntent orchestrates both LLM and non-LLM agents, supporting dynamic re-prompting and contextual feedback to robustly execute user-defined intents with minimal human intervention.
Our implementation of NetIntent across both ODL and ONOS SDN controllers achieves a consistent and adaptive end-to-end IBN realization.

\end{abstract}


\maketitle

\section{INTRODUCTION}

Software-Defined Networking (SDN) was emerged as a revolutionary paradigm to address the challenges of traditional networks~\cite{kreutz2015software, park2023technology}. SDN decouples the network's control functions from its data forwarding functions and enables centralized control and programmability, which are instrumental in simplifying and automating network operations. 
SDN has become a cornerstone for building agile, efficient networks that can adapt to future needs. 
Several SDN deployments, such as Google's B4~\cite{hong2018b4} and Microsoft's SWAN~\cite{hong2013achieving}, demonstrate real-world scalability.
However, SDN's promise of automation is not without limitations. Although it reduces the need for manual configuration at the device level, human intervention is still required to translate business goals or high-level policies into instructions for the SDN controller to realize them on the data plane. OpenDaylight (ODL)~\cite{opendaylight2015} and Open Network Operating System (ONOS)~\cite{ONOS_2024} are two commonly used industry-grade SDN controllers. Organizations like AT\&T and Orange extensively use both ODL and ONOS~\cite{siliconangle2015attodl,orangeopensourceSDN, cmc.2024.047009}. ODL uses DLUX and ONOS uses web GUI to enable users to configure policies through a user-friendly interface~\cite{odl_docs_dlux}. 
However, this process remains time-consuming and prone to human error, especially during initial setup and for complex network configurations.

Intent-Based Networking (IBN) emerges as a natural evolution of SDN, with the aim of simplifying network management by bridging the gap between high-level policy objectives and low-level network configurations~\cite{zeydan2020recent, leivadeas2022survey, rfc9315}.
Industry leaders such as Cisco and Nokia have been instrumental in developing and popularizing IBN in their data centers~\cite{szigeti2018cisco, nokia2024doc}.
IBN enables operators to define what the network should achieve using high-level, human-readable \textit{intents}, rather than specifying how the network should be configured. Formally, intent is defined as a set of operational goals that a network is supposed to meet and outcomes that a network is supposed to deliver, expressed declaratively without specifying how to implement them~\cite{rfc9315}. Intents such as prioritizing video traffic over other types of traffic in an ISP network, or ensuring latency below 5 ms with a minimum bandwidth allocation of 50 Mbps for the URLLC slice in a 5G core network, express desired outcomes without specifying implementation details. This declarative approach enables greater automation and dynamic optimization across different types of networks. Thus, IBN elevates SDN from being a merely programming platform to an autonomous goal-driven architecture.

While IBN significantly reduces the complexity of network management, current implementations still often require operators to express these intents in structured formats like NSD, JSON, XML, or YAML\cite{leivadeas2022survey}. Translating natural-language intents into these machine-readable formats imposes a technical barrier, as it demands familiarity with underlying data models (e.g., YANG \cite{opendaylight_openflowplugin}) and their associated syntax and semantics. For instance, installing flow rules via an ODL controller requires the operator to craft a JSON message aligned with a specific YANG schema, which is an error-prone task that increases operational overhead and risk of misconfiguration.

This gap between high-level intent expression and low-level implementation limits the full potential of IBN. To truly realize the vision of intent-driven networking in which users can express goals in natural language and delegate implementation to the system, there is a need for intelligent mechanisms that can parse, interpret, and compile intents with minimal human intervention. To unlock the full potential of IBN, the network system must support three critical functions: intent translation, intent activation, and intent assurance, to ensure seamless validation and automation of user-defined intents. These functions, when described sequentially, are known as intent lifecycle~\cite{leivadeas2022survey}.

Recent research efforts used large language models (LLMs)~\cite{liu2024large, 10829820, zhao2024explainability} to enhance network management through IBN. These studies mainly focused on intent translation and did not adequately address the challenges of conflicting intents as well as intent assurance.
Further, autonomous end-to-end orchestration of intent lifecycle is still under-addressed. For example, previous work~\cite{10574890,han2025network,mekrache2024llm,wang2024netconfeval,fuad2024intent,tu2025intent,li2024preconfig} did not systematically evaluate the performance of LLMs throughout the entire intent management lifecycle, particularly in critical aspects such as conflict detection and intent assurance. In addition, these works do not cater to SDN controllers like ODL and ONOS without a major modification to their proposed system. Moreover, existing research often relies on closed-source models like ChatGPT, which limit the applicability and transparency of the real world.

To systematically assess the capabilities of LLMs in automating end-to-end IBN tasks, we introduce \textit{\textbf{IBNBench}} to perform a comprehensive benchmarking study that includes intent translation and conflict detection. Existing efforts in this domain typically evaluate few LLMs and focus on limited datasets. In contrast, we benchmark 33 open-source LLMs spanning a diverse range of model sizes and architectures on six datasets, including two existing ones and four newly proposed benchmarks by us. These include the \textit{Intent2Flow-ODL} and \textit{Intent2Flow-ONOS} datasets, which, to the best of our knowledge, are the first to represent natural language intents paired with controller-specific flow rule configurations for both ODL and ONOS. In addition, our \textit{FlowConflict-ODL} and \textit{FlowConflict-ONOS} datasets provide the first curated examples of annotated flow rule pairs for benchmarking LLMs on conflict detection. These datasets enable structured and reproducible comparisons across models and serve as practical testbeds for real-world SDN scenarios. By releasing these datasets and benchmarking results, we provide the research community with essential tools and baselines to further advance LLM-driven IBN systems. To the best of our knowledge, no prior work has conducted such an extensive benchmarking effort across this combination of \textit{tasks, models, and datasets}.

While IBNBench reveals the capabilities and limitations of current LLMs in handling core IBN tasks such as translation and conflict detection, the broader question of how LLMs can be harnessed for end-to-end IBN realization remains unexplored. Previous studies predominantly addressed intent translation, often overlooking critical challenges such as conflict detection, closed-loop assurance, and controller-specific adaptability. To address these gaps, we develop \textit{\textbf{NetIntent}}, an end-to-end intent lifecycle architecture that encompasses intent translation, activation, and assurance, specifically targeting real-world applications for industry-grade SDN controllers, ODL, and ONOS. 

Although modern networks encompass a vast and evolving landscape of thousands of operational intents and configuration tasks, in this paper, we focus specifically on three foundational categories of network intents: \emph{Forwarding} (e.g., routing traffic to specific ports or interfaces), \emph{Security} (e.g., blocking or dropping traffic), and \emph{QoS (Quality of service)} (e.g., prioritizing specific traffic using queues). Our objective is to demonstrate how LLMs can be effectively integrated within an IBN system to enable intent translation, activation, and assurance in a structured and explainable manner. By focusing on a representative subset of fundamental intents, we provide a practical and reproducible system that serves as a proof-of-concept. This example can act as a template for future research aimed at extending LLM-based IBN systems to support broader and more intents in a complex operational environments.

\begin{figure}[t!]
    \centering
    \includegraphics[width=\columnwidth]{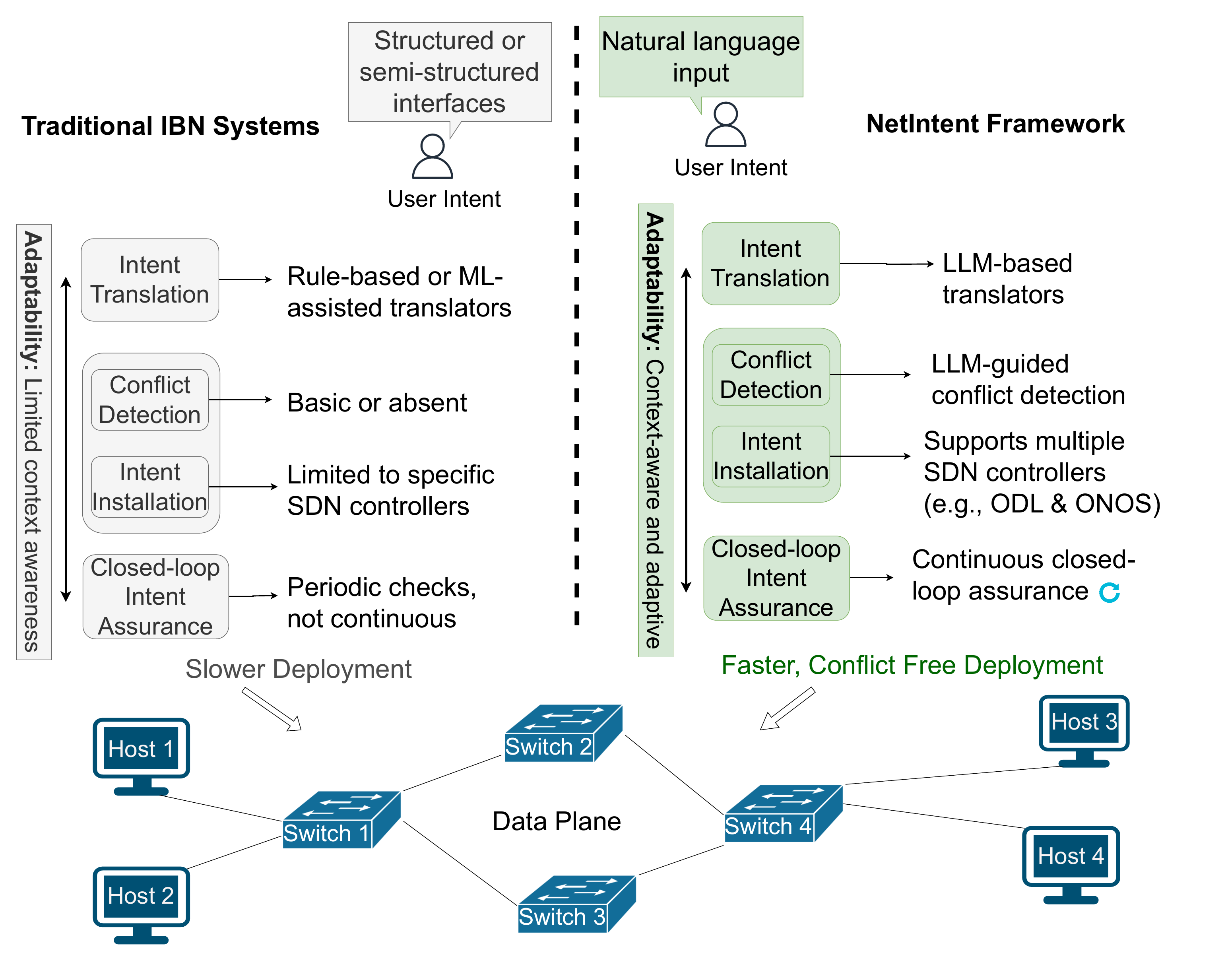}
    \caption{Illustration of NetIntent framework; comparison with prior IBN systems in terms of intent life-cycle.}
    \label{fig:contribution}
\end{figure}

In Fig. \ref{fig:contribution}, we present a comparative overview of NetIntent and traditional IBN systems throughout the entire intent processing workflow. On the left side of Fig. \ref{fig:contribution}, conventional IBN systems rely on structured or semi-structured user interfaces to capture operator intent. These systems typically execute a linear and static sequence that includes intent translation, conflict detection, flow rule installation, and optional closed-loop assurance. However, their adaptability is limited, as they often lack contextual reasoning and dynamic reactivity, resulting in slower deployments and increased likelihood of misconfigurations. In contrast, NetIntent, depicted on the right side of Fig.~\ref{fig:contribution}, enables users to express intents directly in natural language. It employs LLMs in a context-aware and adaptive manner which enables it to dynamically update LLM-prompts based on feedback and failed assurance checks. This design enables NetIntent to achieve faster and conflict-free intent deployment by proactively identifying issues, minimizing the need for manual intervention.

\begin{table*}[t!]
\centering
\caption{Comparison of NetIntent with relevant works on IBN}
\footnotesize
\resizebox{\textwidth}{!}{%
\begin{tabular}{llcccccccccccccccc}
\toprule
\textbf{Category} & \textbf{Aspect} & 
\textbf{\cite{jacobs2019deploying}} & \textbf{\cite{10979966}} & \textbf{\cite{guo2023configreco}} & \textbf{\cite{alsudais2017hey}} & \textbf{\cite{mahtout2020using}} & \textbf{\cite{yang2020intent}} & \textbf{\cite{kiran2018enabling}} & 
\textbf{\cite{han2025network}} & \textbf{\cite{mekrache2024llm}} & \textbf{\cite{wang2024netconfeval}} & \textbf{\cite{fuad2024intent}} & \textbf{\cite{tu2025intent}} & \textbf{\cite{10574890}} & \textbf{\cite{li2024preconfig}} & \textbf{\cite{dzeparoska2024intent}} &
\textbf{NetIntent} \\
\midrule
\rowcolor{gray!10}
\multicolumn{16}{l}{\textbf{Methodology}} \\
& LLM-based   & \xmark & \xmark & \xmark & \xmark & \xmark & \xmark & \xmark & \cmark & \cmark & \cmark & \cmark & \cmark & \cmark & \cmark & \cmark & \cmark \\
& Non-LLM     & \cmark & \cmark & \cmark & \cmark & \cmark & \cmark & \cmark & \xmark & \xmark & \xmark & \xmark & \xmark & \xmark & \xmark & \xmark & \xmark \\
\midrule
\rowcolor{gray!10}
\multicolumn{16}{l}{\textbf{Implementation}} \\
& Automated   & \cmark & \cmark & \xmark & \xmark & \xmark & \cmark & \cmark & \cmark & \cmark & \xmark & \xmark & \cmark & \cmark & \cmark & \cmark & \cmark \\
\midrule
\rowcolor{gray!10}
\multicolumn{16}{l}{\textbf{Input Method}} \\
& Natural Language Input & \cmark & \xmark & \xmark & \cmark & \cmark & \cmark & \cmark & \cmark & \cmark & \cmark & \cmark & \cmark & \cmark & \cmark & \cmark & \cmark \\
\midrule
\rowcolor{gray!10}
\multicolumn{16}{l}{\textbf{IBN Lifecycle Functions}} \\
& Intent Translation         & \cmark & \xmark & \cmark & \cmark & \cmark & \cmark & \cmark & \cmark & \cmark & \cmark & \cmark & \cmark & \cmark & \cmark & \cmark & \cmark \\
& Conflict Detection         & \cmark & \xmark & \xmark & \xmark & \xmark & \xmark & \xmark & \xmark & \xmark & \cmark & \xmark & \xmark & \xmark & \xmark & \xmark & \cmark \\
& Activation/Deployment      & \cmark & \cmark & \xmark & \xmark & \cmark & \cmark & \cmark & \cmark & \xmark & \cmark & \cmark & \xmark & \xmark & \cmark & \cmark & \cmark \\
& Intent Assurance           & \xmark & \cmark & \xmark & \xmark & \xmark & \cmark & \xmark & \cmark & \xmark & \xmark & \xmark & \xmark & \xmark & \xmark & \cmark & \cmark \\
\midrule
\rowcolor{gray!10}
\multicolumn{16}{l}{\textbf{LLM Application}} \\
& \# LLMs Evaluated          & -- & -- & -- & -- & -- & -- & -- & 1 & 3 & at least 7 & 8 & 7 & 3 & at least 3 & at least 1 & at least 33 \\
& LLM Type                   & -- & -- & -- & -- & -- & -- & -- & open-src & open-src & mixed & mixed & mixed & open-src & mixed & closed-src & open-src \\
\bottomrule
\end{tabular}
}
\label{tab:llm-non-llm-ibn}
\end{table*}

Table~\ref{tab:llm-non-llm-ibn} presents a comparative overview of the most relevant work on IBN, categorized by key attributes of the system. The methodology category distinguishes between LLM-based approaches, where at least one LLM is employed at any stage of the system pipeline (e.g., intent parsing, translation, deployment) as seen in works~\cite{han2025network,mekrache2024llm,wang2024netconfeval,fuad2024intent,tu2025intent,10574890,li2024preconfig, dzeparoska2024intent}—and Non-LLM systems~\cite{jacobs2019deploying,10979966,guo2023configreco,alsudais2017hey,mahtout2020using,yang2020intent,kiran2018enabling} that rely on traditional programmatic logic or conventional machine learning. The Implementation category indicates whether the system is fully automated (no human input required beyond initial intent, except in rare exceptions), which applies to works like~\cite{han2025network,mekrache2024llm,tu2025intent,10574890,jacobs2019deploying,yang2020intent,kiran2018enabling}, or semi-automated (requiring human assistance for prompting or execution), as in~\cite{wang2024netconfeval,fuad2024intent,guo2023configreco,alsudais2017hey,mahtout2020using}. Natural language input highlights whether the system can parse unstructured user intents, and IBN Lifecycle Functions include capabilities such as intent translation, conflict detection, deployment, and assurance. Finally, the LLM Application category reflects the extent and type of LLM usage, where applicable. While these works demonstrate early promise in automating various IBN tasks, our understanding of how open-source LLM-based systems can autonomously manage the full IBN lifecycle—from natural language input to assured deployment—remains limited.

In the following points, we summarize our contributions.

\begin{itemize}
\item We perform a benchmarking of 33 open-source LLMs using new and existing datasets comprising natural language intents and their corresponding JSON configurations. This evaluation assesses LLM performance in both intent translation and conflict detection, offering insights into their suitability for IBN tasks.
\item We introduce IBNBench, a suit of four new datasets, \textit{Intent2Flow-ODL}, \textit{Intent2Flow-ONOS}, \textit{FlowConflict-ODL}, and \textit{FlowConflict-ONOS}, make them open source to support reproducible research. Intent2Flow datasets consist of natural language configuration intents paired with structured flow rules, while the FlowConflict datasets contain flow rule pairs annotated for conflict detection. These resources serve as practical benchmarks for evaluating LLM-driven IBN systems.
\item We propose \textit{NetIntent}, a novel end-to-end architecture for intent-based networking that leverages LLMs to automate the full intent lifecycle, including translation, activation, and assurance.
\item To demonstrate the generalizability and practical applicability of our approach, we implement and evaluate NetIntent on two widely-used SDN controllers: ODL and ONOS.

\end{itemize}


\vspace{-0.3cm}
\section{BACKGROUND AND RELATED WORKS}
\label{sec:background_and_related_work}
The lifecycle of an intent in IBN encompasses distinct phases, namely Intent Translation, Intent Activation, and Intent Assurance. Each phase involves specific tasks and challenges essential to effectively operationalize high-level network objectives.

\vspace{-0.3cm}
\subsection{INTENT TRANSLATION}
Intent translation is the process of converting user-defined high-level goals into precise, low-level network configurations that can be executed by controllers. This step is fundamental to IBN and its success determines how accurately the network enforces operator intentions. Key challenges in this phase include managing the inherent ambiguity of natural language, refining hierarchical or multi-layered intents, resolving multiple valid configurations, and supporting controller- or vendor-specific implementations. Traditional NLP techniques, while useful for rule-based intent parsing~\cite{10175400, jacobs2019deploying,guo2023configreco,alsudais2017hey,mahtout2020using,yang2020intent,kiran2018enabling}, often fall short due to the linguistic variability and context dependence of human input in natural language~\cite{creanga-dinu-2024-designing}.

Recent works~\cite{han2025network,mekrache2024llm,wang2024netconfeval,fuad2024intent,tu2025intent,10574890,li2024preconfig, dzeparoska2024intent} have leveraged LLMs to improve intent translation accuracy. For example, NetConfEval~\cite{wang2024netconfeval} evaluates ChatGPT and Codellama on translating formal network specifications—such as reachability, waypoints, and load balancing—into structured JSON rules. Their dataset includes intents like \textit{``Traffic originating from Istanbul can reach the subnet 100.0.9.0/24 via Rotterdam, using two paths"}. The work in~\cite{tu2025intent} introduces another benchmark focused on NFV (Network Function Virtualization) configuration, covering service function chaining and resource allocation. Both studies use in-context learning\cite{dong2024survey}, where example input-output pairs are embedded in the LLM prompt to guide translation. The NFV configuration work also explores a continuous learning setup, where previous corrections are reused to improve future translations.

Despite these advances, several limitations persist. First, many existing approaches rely on closed-source models such as ChatGPT~\cite{wang2024netconfeval,fuad2024intent}, which limit reproducibility due to closed-source nature, raise privacy concerns due to lack of transparency, and hinder real-world deployment due to the cost. Second, while previous work benchmarks LLMs for translation, they typically evaluate only a handful of models, for example, 3 in~\cite{mekrache2024llm}, 6 in~\cite{tu2025intent}, or 7 in~\cite{wang2024netconfeval}—and restrict evaluation to one or two datasets. Third, these studies focus only on abstract configuration goals (e.g., formal policies or NFV rules), overlooking practical translation targets like SDN controller-specific formats. Finally, there remains a critical gap in understanding how LLMs perform across a wider range of IBN tasks, particularly when the target output must conform to real-world SDN controller schemas like ODL or ONOS.

\vspace{-0.3cm}
\subsection{INTENT ACTIVATION}
\label{subsec:background_intent_activation}
Intent activation refers to the deployment of translated intents onto the network infrastructure. This phase encompasses two critical tasks: detecting and resolving conflicts between new and existing configurations, and correctly installing the validated configuration on the appropriate network devices. 

Conflicts can arise from multiple overlapping intents that often issued by different users or systems and target the same network scope with contradictory goals. These conflicts may stem from policy misalignment, intent ambiguity, or resource contention. 

\renewcommand{\arraystretch}{0.6}
\begin{table*}[t!]
\caption{Flow rule table}
\label{tab:conflict_flow_rules}
\centering
\small
\resizebox{\textwidth}{!}{%
\begin{tabular}{lccccccccc}
\toprule
\textbf{Rule \#} & \textbf{Priority} & \textbf{Source MAC} & \textbf{Dest MAC} & \textbf{Source IP} & \textbf{Dest IP} & \textbf{Protocol} & \textbf{Source Port} & \textbf{Dest Port} & \textbf{Action} \\
\midrule
1 & 61 & * & * & 192.168.10.0/24 & 172.16.1.100 & tcp & * & * & forward \\
2 & 60 & * & * & 192.168.10.25 & 172.16.1.100 & tcp & * & 443 & forward \\ 3 & 62 & * & * & 192.168.10.25 & 172.16.1.0/24 & tcp & * & * & forward \\
4 & 63 & * & * & 192.168.10.0/24 & 172.16.1.100 & tcp & * & * & drop \\
5 & 64 & * & * & 192.168.10.25 & 172.16.1.100 & tcp & * & * & drop \\
6 & 61 & * & * & 192.168.0.0/16 & 172.16.1.100 & tcp & * & * & drop \\
7 & 65 & * & * & 192.168.10.25 & 172.16.1.0/24 & tcp & * & 4000--4010 & drop \\
8 & 67 & aa:bb:cc:dd:ee:01 & ff:ee:dd:cc:bb:aa & * & * & * & * & * & forward \\
9 & 68 & * & * & * & * & tcp & * & 443 & drop \\
\bottomrule
\end{tabular}%
}
\end{table*}

Existing works \cite{pisharody2017brew, khairi2021detection} described six types of conflicts in SDN environments and developed methods to identify them. In Table \ref{tab:conflict_flow_rules}, we present 9 flow rules. This table is adapted from \cite{pisharody2017brew} to demonstrate different types of conflicts. A redundancy conflict occurs when a specific rule (Rule 2) is fully covered by a more general one (Rule 1) with the same action. Shadowing conflict is observed when a more general rule (Rule 4) with higher priority overrides a more specific rule (Rule 1) with a different action. Generalization conflict happens when a specific rule (Rule 5) with higher priority conflicts in action with a broader rule (Rule 1). Correlation conflict is identified between Rule 6 and Rule 1, where their match spaces partially overlap without a subset relationship and their actions differ. Overlap conflict, as seen between Rule 6 and Rule 7, involves intersecting address spaces with the same action. Lastly, Rule 4 and Rule 8 illustrate imbrication conflict, where rules overlap on one protocol layer (e.g., MAC) but not on another (e.g., IP), causing cross-layer ambiguity.

Traditional conflict resolution techniques—such as logical policy evaluation, verification against the network state, or graph-based mapping~\cite{zhang2021conflict, cui2018transaction, leivadeas2022survey,zeydan2020recent}. For example, VeriFlow \cite{khurshid2012veriflow} and NetPlumber \cite{kazemian2013real} enable real-time conflict detection in SDN by analyzing flow updates for violations such as loops or black holes, with support for basic automated resolution like blocking or flagging rules. However, they lack support for complex or non-flow-based intent conflicts. In contrast, Batfish \cite{fogel2015general} performs offline static analysis to detect configuration errors with detailed provenance but does not support real-time detection or automated resolution, making it less suitable for dynamic IBN scenarios. Furthermore, current approaches largely lack mechanism to explain why a conflict occurred, limiting their ability to clearly indicate underlying causes of conflicts or recommend actionable resolution strategies, thus highlighting the need for enhanced intent resolution methods. While some recent works have begun to explore conflict handling using LLMs, their scope remains narrow. For example, NetConfEval~\cite{wang2024netconfeval} demonstrates basic conflict detection using LLMs by identifying mutually exclusive reachability goals. However, to extent to which LLM can do conflict detection for SDN environments remains underexplored.

\vspace{-0.3cm}
\subsection{INTENT ASSURANCE}
Intent assurance represents the final and most iterative phase of the IBN lifecycle, responsible for continuously verifying that the operational network state aligns with the originally expressed user intents~\cite{10230112}. This involves not only validating the correctness of intent translation and deployment but also ensuring ongoing enforcement through real-time monitoring, predictive analytics, and intelligent feedback mechanisms. A critical challenge in this phase is managing intent drift \cite{rfc9315}, a condition where network behavior gradually deviates from the intended objectives due to dynamic changes in traffic patterns, topology, or device states. Effective intent assurance must detect such drift and trigger corrective actions to restore compliance and maintain intent fidelity over time.

Traditional approaches to intent assurance typically fall into static and dynamic categories~\cite{leivadeas2022survey}. Static methods verify if the intended configurations exist on the correct device and match the expected structure. For instance, in ODL, there are configurational data store and operational data store. Configurational data store reflects all installed flow rules (active or passive) while the operational data store reflects the active flow rules. 
For static assurance, the operational data store can be used to determine the status of flow rules. For example, the operational data store in ODL is queried to inspect whether the \texttt{packet-count} field in the \texttt{flow-statistics} section of a flow rule contains a non-zero value. A non-zero count indicates that the rule has matched traffic, suggesting packet drops if the rule is intended to drop packets, or forwarding if it is meant to forward them. However, static assurance often fails to detect runtime violations.

In contrast, dynamic assurance mechanisms rely on continuous monitoring of key performance indicators (KPIs) such as delay, throughput, packet loss, and queue utilization to ensure that the deployed flows exhibit the desired behavior. For example, VeriFlow \cite{khurshid2012veriflow} and NetPlumber \cite{kazemian2013real} provide real-time flow rule assurance in SDN by incrementally checking rule compliance with network invariants or policies during updates, offering fast but limited verification focused on flow-level behavior. Batfish \cite{fogel2015general}, on the other hand, performs static assurance by analyzing configurations pre-deployment using constraint solving, which enables thorough checks but lacks the real-time responsiveness needed for dynamic IBN scenarios.

LLM-based approaches have only recently begun to touch on intent assurance. Although the work in~\cite{10574890} is implemented on a 5G testbed and introduces a modular LLM-centric framework to cover the full intent lifecycle, there is no evidence on how and to what extent intent assurance can be realized. 
Other works such as~\cite{10272352} explore fault localization using LLMs but do not provide a generalized assurance framework. While most existing LLM-based IBN systems~\cite{mekrache2024llm,wang2024netconfeval,fuad2024intent,tu2025intent,10574890,li2024preconfig} do not implement assurance in a closed-loop or controller-aware manner, the work in \cite{dzeparoska2024intent} is among the earliest to propose an LLM-driven assurance system capable of detecting intent performance drift and triggering corrective actions. However, its reliance on the closed-source LLM ChatGPT may limit transparency, reproducibility, and broader adoption.

\vspace{-0.2cm}
\section{IBNBench: LLM-BASED IBN EVALUATION BENCHMARK}
\subsection{SELECTION OF LLMs}
To ensure a focused yet practical evaluation, we selected 33 open-source LLMs, as in Table \ref{tab:llm_model_list}, based on a combination of resource feasibility, task relevance, and community adoption. Due to computational and time constraints, we limited the scope to a representative subset of models that are well-suited for intent translation and conflict detection—excluding models designed for unrelated domains. Additionally, we prioritized LLMs that are popular within the research community, frequently cited, and actively maintained. All selected models are readily deployable, making them suitable candidates for reproducible benchmarking and real-world integration in IBN systems. These LLMs vary in the number of parameters, which directly influence their language processing and generation capabilities. Larger models with more parameters generally exhibit better understanding and nuanced responses but at the cost of increased memory requirements, higher energy consumption and slower processing speeds.

\renewcommand{\arraystretch}{0.6}
\begin{table}[t!]
\caption{List of LLMs used for benchmarking (grouped by parameter size in billions)}
\label{tab:llm_model_list}
\centering
\normalsize
\resizebox{\columnwidth}{!}{%
\begin{tabular}{p{1cm} >{\raggedright\setstretch{0.5}\arraybackslash}p{9.6cm}}
\toprule
\textbf{Parameter Size} & \hspace{2em} \textbf{Models} \\
\midrule
1–3B & Starcoder:3b, Llama3.2:3b, Phi3:3.8b, Orca-mini:3b, Starcoder2:3b, TinyLlama:1.1b, Deepseek-coder:1.3b, Phi:2.7b \\

4–6B & Qwen:4b, Yi:6b \\

7–9B & Codellama:7b, Llama2:7b, Llama3:8b, Llama3.1:8b, Qwen2.5:7b, Openchat:7b, Marco-o1:7b, Mistral:7b, Dolphin-Mistral:7b, Wizardlm2:7b, Codegemma:7b, Zephyr:7b, Llava-Llama3:8b, Qwen2:7b \\

10–20B & Mistral-nemo:12b, Deepseek-coder-v2:16b \\

20–30B & Gemma2:27b, Codestral:22b \\

30B+ & QwQ-abliterated:32b, QwQ-fusion:32b, QwQ:32b, Codellama:34b, Command-r:35b \\
\bottomrule
\end{tabular}%
}
\end{table}

\subsection{SELECTION OF DATASETS}
\label{subsec:benchmark_datasets}
We benchmarked the LLMs using six datasets—four newly proposed by us and two existing ones~\cite{wang2024netconfeval, tu2025intent}. The creation process of the proposed datasets is detailed in Sec. \ref{dataset_preparation}, and a summary of all datasets is provided in Table~\ref{tab:benchmark_datasets}.

\renewcommand{\arraystretch}{1}
\begin{table}[t!]
\caption{Summary of datasets used for LLM benchmarking; four newly proposed by us and two existing ones}
\label{tab:benchmark_datasets}
\centering
\normalsize
\resizebox{\columnwidth}{!}{%
\begin{tabular}{lccccc}
\toprule
\textbf{Dataset} & \textbf{Proposed} & \textbf{Task} & \textbf{Samples} & \textbf{Conflict Pairs} & \textbf{Evaluation Metric} \\
\midrule
Intent2Flow-ODL & Yes & Translation & 52 & -- & Semantic Accuracy, Runtime \\
Intent2Flow-ONOS & Yes & Translation & 50 & -- & Semantic Accuracy, Runtime \\
Formal Spec.~\cite{wang2024netconfeval} & No & Translation & 1500 & -- & Field Presence Accuracy, Runtime \\
NFV Config.~\cite{tu2025intent} & No & Translation & 120 & -- & Exact Structural Match, Runtime \\
FlowConflict-ODL & Yes & Conflict Detection & 50 & 4 & TP, TN, FP, FN, Runtime \\
FlowConflict-ONOS & Yes & Conflict Detection & 62 & 10 & TP, TN, FP, FN, Runtime \\
\bottomrule
\end{tabular}%
}
\end{table}

The LLMs are benchmarked for two different IBN tasks: intent translation (from natural language intent to JSON structured flow rules) and conflict detection. The benchmarking for intent translation is done using four different datasets. Among them, two belong to the proposed IBNBenh (Intent2Flow-ODL and Intent2Flow-ONOS) mentioned in Sec. \ref{dataset_preparation}, the third one is the Formal specification dataset ~\cite{wang2024netconfeval} and the fourth one is the NFV configuration dataset ~\cite{tu2025intent}. The reason to choose the datasets ~\cite{wang2024netconfeval} and ~\cite{tu2025intent} is that they contain natural language intent and corresponding JSON formatted translation. The Formal specification dataset translates natural language intents into JSON structure for specifically three type of requirements: reachability, way-points and load balancing, while the NFV configuration dataset include natural language intent and corresponding JSON formatted NFV configuration. However, these datasets are different from our proposed datasets. Our datasets specifically target ODL and ONOS SDN controllers and contains actual flow rules that were tested and verified. Hence, they can be readily used to benchmark any LLM for ODL or ONOS SDN controller application.

As for the benchmarking the LLMs for conflict detection task, we use our proposed IBNBench's FlowConflict-ODL and FlowConflict-ONOS datasets. We do not include the Formal specification JSON and NFV configuration JSON. The reasons is that we detect conflict based on JSON structured configuration and all the datasets use JSON formatted configuration. Hence, it is sufficient to evaluate the LLMs on the ODL and ONOS flow rules as they represent well structured JSON configuration found in other datasets.

As for the sample size of the datasets, the NFV configuration dataset contains 120 pairs of samples, the Formal specification dataset comprises 1500 sample pairs. Our proposed datasets Intent2Flow-ODL and Intent2Flow-ONOS include 52 and 50 pairs of samples receptively, while the FlowConflict-ODL and FlowConflict-ONOS datasets include 60 and 74 pairs of samples receptively. 

For intent translation, each dataset was split, with 50\% used as the source of context examples and the remaining 50\% as the source of test cases. A context example (context example) in the LLM context is an input-output pair provided in a prompt to guide the model's behavior in tasks like few-shot learning, demonstrating the desired format and content for the output. For conflict detection, 62 pairs of flow rules were selected from the FlowConflict-ONOS dataset, including 10 pairs with conflicts. From the FlowConflict-ODL dataset, 50 pairs were selected, of which 4 contained conflicts.

\vspace{-0.3cm}
\subsection{BENCHMARK METRICS}
For evaluating the performance of LLMs in the intent translation task, we use accuracy as the primary metric. Besides we report the running time. The running time includes the time from the submission of the query to the LLM until the LLM produces the output, including the time for dynamic selection of context examples. 

The method of computing accuracy varies across the four datasets, depending on the nature of their expected outputs. For the proposed Intent2Flow-ODL and Intent2Flow-ONOS datasets, a translation is considered correct if the generated JSON is semantically equivalent to the expected JSON—allowing for differences in field ordering, formatting, or numeric representation. In contrast, the Formal Specification dataset uses a more relaxed comparison: it considers a translation correct if key expected intent fields (e.g., reachability, waypoint, load balancing) are present in the result, without requiring full structural matching. For the NFV Configuration dataset, a stricter approach is adopted where accuracy is computed by checking for exact structural equality after recursively sorting dictionary keys and list elements.

As for evaluating the performance of LLMs in conflict detection task, we set one conflict as defined in Sec. II-\ref{subsec:background_intent_activation} for the flow rule pairs given to an LLM. We record true positive, true negative, false positive, false negative for each LLM by comparing against the ground truth. Besides, we report the time duration from the submission of the prompt to the LLM until the LLM produces an output. In the benchmarking process, we ask LLMs to identifying potential conflicts between flow rules which requires examining whether multiple rules overlap in their matching criteria while prescribing different output actions. For example, two rules that match the same type of traffic, such as TCP packets on a specific port from the same input port, and apply different output ports or actions like drop versus forward, are considered conflicting. This is evident when two flow entries match the same traffic conditions but direct it differently, creating ambiguity in packet handling. Conflict detection, therefore, involves comparing match fields and examining whether their actions diverge for overlapping traffic. 

\vspace{-0.3cm}
\subsection{LIMITATIONS}
The evaluation focuses on three categories of intent: forwarding, security, and QoS. Broader categories of network intents, such as service function chaining or dynamic path optimization, are not currently evaluated. However, the benchmarking framework is designed to be extensible. New intent categories and flow rule patterns can be integrated in IBNBench, and we plan to expand it over time while enabling community contributions to promote broader coverage and reproducibility.

\section{DEVELOPING IBNBench}
\label{dataset_preparation}

To address the gap in evaluating LLM-based systems on intent translation and detection of conflicting flow rules, we designed IBNBench, a set of four datasets, two for intent translation and two for conflict detection. The datasets for translating natural langugae intents into JSON formatted flow rules are named Intent2Flow-ODL and Intent2Flow-ONOS, based on the target SDN controller. As for the datasets developed for conflict detection, they are named FlowConflict-ODL and FlowConflict-ONOS according to the target SDN controller.

\vspace{-0.3cm}
\subsection{Intent2Flow-ODL AND Intent2Flow-ONOS DATASETS}
Each of the datasets, Intent2Flow-ODL, Intent2Flow-ONOS, consist of 50 pairs of real-world-inspired network intents in natural language and their JOSN formatted translation. Each intent specifies high-level operational goals such as routing, blocking, or prioritizing traffic within a programmable network environment. We consider that the user intents belong to three primary categories: \emph{Forwarding}, \emph{Security}, and \emph{QoS} based on the semantics of their described action. Intents that included explicit blocking, dropping, or denying of traffic are under \emph{Security}, while intents specifying packet forwarding without prioritization are under \emph{Forwarding}. Intents specifying queue assignments, traffic prioritization, or latency/bandwidth guarantees are \emph{QoS} intents. The datasets contain 22 forwarding intents, 10 security intents and 18 QoS intents.

To construct the datasets, we adopted both manual and LLM-generated content. A diamond topology was created using Mininet~\cite{dholakiya2021survey}  for both ODL and ONOS, consisting of four OpenFlow switches and four hosts, as shown in Fig. \ref{topology}. Based on this network, we crafted intents and produced their corresponding ODL/ONOS flow rules in JSON format. LLMs were leveraged to generate supplementary intents with variation in linguistic expression as well as corresponding JSON flow rules. Before inclusion in the datasets, these LLM-generated flow rules were manually corrected and deployed in switches to verify their effectiveness. We provide the full datasets on GitHub~\cite{NetIntent2025}. In Table \ref{tab:example_intents}, some examples of intents are shown.

\renewcommand{\arraystretch}{0.6}
\begin{table}[t!]
\caption{Example intents from IBNBench}
\label{tab:example_intents}
\centering
\normalsize
\resizebox{\columnwidth}{!}{%
\begin{tabular}{p{1.6cm} >{\setstretch{0.5}}p{9.5cm}}
\toprule
\textbf{Intent Type} & \textbf{Example Intent} \\
\midrule
\multirow{2}{*}{Forwarding} 
& \textbullet\ Forward traffic entering on port 1 of switch 2 to port 2. \\
& \textbullet\ In switch 1, forward all TCP traffic not matching higher-priority rules through port 1. \\
\midrule
\multirow{2}{*}{Security} 
& \textbullet\ In switch 4, block all IPv4 traffic from 10.0.0.1 to 10.0.0.4 with a high priority, ensuring the switch operates as a firewall. \\
& \textbullet\ Drop all packets with a source IP of 10.0.0.1 and destination IP of 10.0.0.4 using node 4. \\
\midrule
\multirow{2}{*}{QoS} 
& \textbullet\ Forward TCP traffic on port 80 destined for 10.0.0.3 via interface 2 of switch 1, assigning it to queue 0 for prioritized handling. \\
& \textbullet\ Route HTTP traffic originating from 192.168.1.2 on port 1 of switch 4 and destined for 10.0.0.5/32 through port 2, ensuring packets are assigned to queue 0 for low-latency and apply VLAN tag 100. \\
\bottomrule
\end{tabular}%
}
\end{table}

\noindent
\textbf{ODL Flow Rules}: flow rule is a structured configuration that defines how network traffic is managed within an SDN environment. In ODL, these rules are determined by pre-defined YANG model \cite{opendaylightOpenFlowPlugin} and typically represented in JSON format. It includes fields such as flow IDs, table IDs, priority, match conditions (e.g., source/destination IP, protocol type), and actions (e.g., forwarding, dropping, or modifying packets). This structured approach allows ODL to efficiently describe and implement complex networking policies, making it suitable for configuring large-scale networks, traffic engineering, and advanced use cases like network slicing or QoS enforcement. In Appendix \ref{appendix:translated_intent}, an example intent and its corresponding ODL flow rule representation is included. More details on ODL flow rules can be found in \cite{opendaylightOpenFlowPlugin}.

\noindent
\textbf{ONOS Flow Rules:} In ONOS, flow rules serve the same purpose as in ODL, defining how packets should be processed based on structured match conditions and corresponding actions. These rules are expressed in JSON format and include key components such as device identifiers, match criteria, priority levels, and treatments. The match field specifies the conditions under which the rule applies, such as Ethernet type, IP protocol, and source or destination IP addresses. The action field, referred to as the treatment, outlines how packets should be handled when a match occurs—for example, forwarding to a specific port, dropping, or assigning traffic to a QoS queue. In Appendix \ref{appendix:translated_intent}, an example intent and its corresponding ONOS flow rule representation is included. More details on ONOS flow rules can be found in \cite{onosprojectFlowRules}.

\vspace{-0.4cm}
\subsection{FlowConflict-ODL AND FlowConflict-ONOS DATASETS}
\label{FlowConflict_datasets}
We used JSON formatted flow rules to create two new datasets to benchmark the LLMs on conflict detection task. They are FlowConflict-ODL and FlowConflict-ONOS datasets. FlowConflict-ODL contains 60 pairs of JSON formatted flow rules designed for ODL SDN controller and FlowConflict-ONOS contains 74 pairs of JSON formatted flow rules designed for ONOS SDN controller. As for the number of conflicting rule pairs, FlowConflict-ODL has 19 pairs of conflicting rules and FlowConflict-ONOS has 27 pairs of conflicting rules. These conflicting rules belong to six conflict category discussed in Sec. II-\ref{subsec:background_intent_activation}. We provide the full datasets on GitHub~\cite{NetIntent2025}.

\vspace{-0.3cm}
\section{NetIntent DESIGN}
\label{prop_arch}

\subsection{PROBLEM MODELING}

Automating the full lifecycle of IBN from high-level natural language intent expression to low-level SDN configuration and validation requires a principled formulation of the underlying decision process. We model this IBN orchestration as a mapping from user intents and network context to controller-executable configurations into the network devices.

Let \( \mathcal{I} \) denote the space of user-defined natural language intents (e.g., QoS policies, security rules, routing goals), and let \( \mathcal{N} \) represent the network context, including topology, installed configurations, and performance metrics. The goal is to synthesize a configuration \( c \in \mathcal{C} \), where \(\mathcal{C} \) is the space of target configurations that can be applied to the SDN controller, such that:

\begin{equation}
\Pi: \mathcal{I} \times \mathcal{N} \rightarrow \mathcal{C}
\end{equation}

This configuration must accurately implement the operator's intent under the current network state. We define a semantic alignment function \( \text{SemMatch}(i, c) \), which returns 1 if the configuration \( c \) satisfies the operational goals embedded in the intent \( i \), and 0 otherwise:

\begin{equation}
\text{SemMatch}(i, c) = 1
\end{equation}

This process must support diverse intent types including forwarding, security, and QoS, across heterogeneous SDN controllers (e.g., ODL, ONOS). The mapping is further constrained by the compliance with the controller-specific schema.

To ensure safe deployment, the generated configuration must be valid and non-conflicting with respect to the current configuration state \( C_{\text{curr}} \) (e.g., overlapping match fields with contradictory actions). This requires satisfying two constraints: (1) schema validity, expressed as \( \text{Valid}(c) = 1 \), and (2) absence of rule-level conflict with existing flow entries, defined as \( \text{Conflict}(c, C_{\text{curr}}) = 0 \).

If a conflict is detected, the system resolves it using a deterministic policy function:

\begin{equation}
\rho: \mathcal{C} \times \mathcal{C} \rightarrow \mathcal{C}
\end{equation}

Here, the first argument to \( \rho \) represents the newly generated candidate configuration, and the second argument represents an existing configuration already deployed in the network. Although both inputs are elements of the same configuration space \( \mathcal{C} \), their roles are distinct in the resolution process. The function \( \rho \) selects which configuration to prioritize, either retaining the new rule or preserving the existing one, based on factors such as intent type (e.g., security over QoS or forwarding), match specificity, and contextual metadata.

After successful installation, the system must verify whether the deployed configuration produces the intended behavior in the live network. This process, known as \textit{intent assurance}, compares the actual operational state \( N_{\text{obs}} \), including real-time telemetry such as packet counters, queue lengths, and latency metrics, associated with the deployed rule \( c \),  against the behavioral expectations implied by the original intent \( i \). An assurance function is defined as:

\begin{equation}
\text{Assure}(i, c, N_{\text{obs}}) =
\begin{cases}
1, & \text{if behavior aligns with intent} \\
0, & \text{otherwise}
\end{cases}
\end{equation}


The ultimate objective is to maximize the number of intents that are semantically realized, safely activated, and successfully assured. Given a batch of intents \( \{i_1, i_2, \dots, i_n\} \), the autonomous orchestration of the intent-based SDN system becomes:

\begin{align}
\max_{\Pi} \sum_{j=1}^{n} \Big[ &\text{SemMatch}(i_j, c_j) \cdot \text{Valid}(c_j) \cdot \nonumber \\
&\neg \text{Conflict}(c_j, C_{\text{curr}}) \cdot \text{Assure}(i_j, c_j, N_{\text{obs}}) \Big]
\end{align} 

This formulation abstracts the IBN pipeline as a composite optimization problem, where the orchestration function \( \Pi \) performs a sequence of reasoning and verification steps. Each step is assisted by LLMs through prompt engineering, conflict resolution, fallback mechanisms, and closed-loop assurance. In the next section, we present our proposed NetIntent system, which operationalizes this formulation to autonomously solve the orchestration problem across the full intent lifecycle.


\vspace{-0.3cm}
\subsection{NetIntent OVERVIEW}

\begin{figure}[t!]
    \centering
    \includegraphics[width=\linewidth]{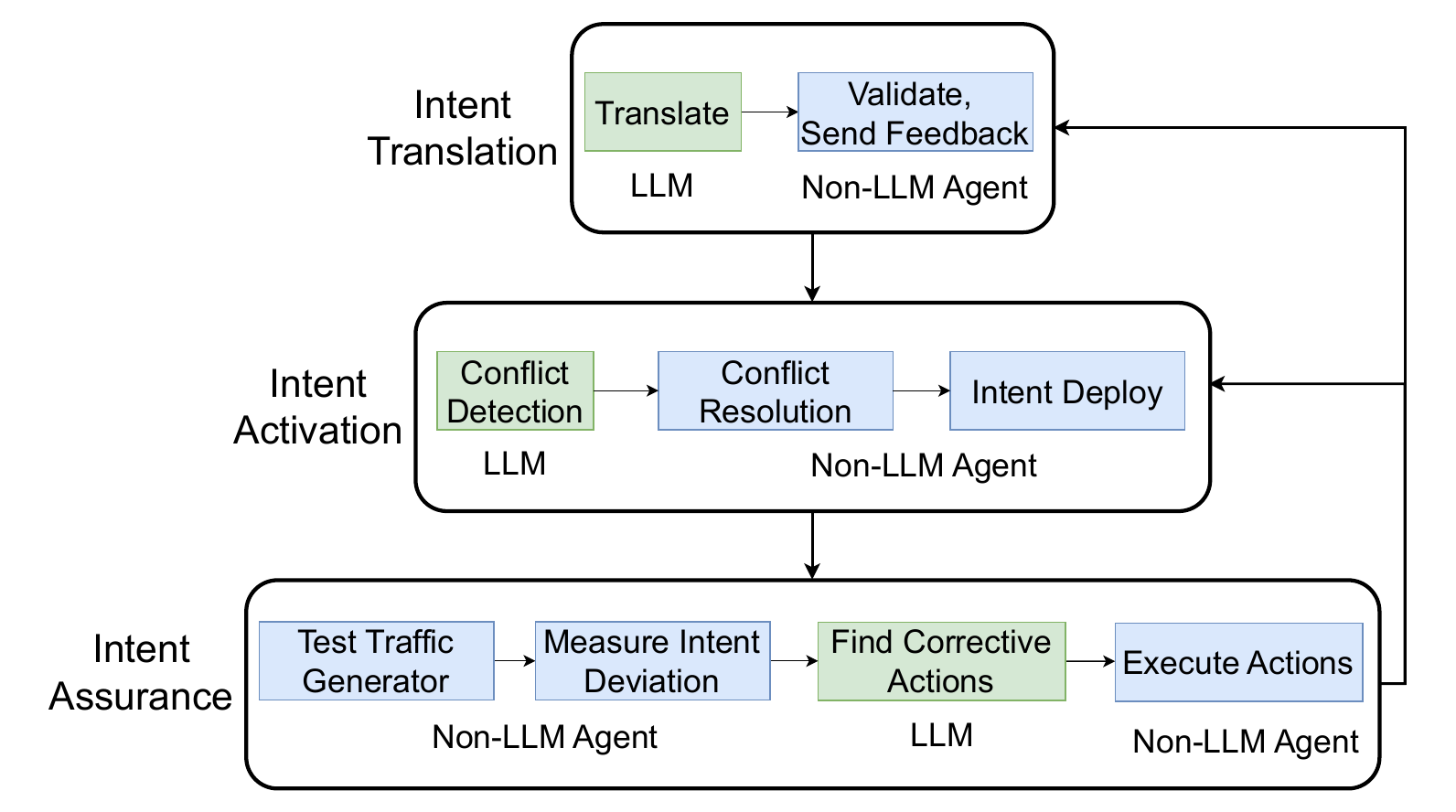}
    \caption{NetIntent overview.}
    \label{architecture_abstract}
\end{figure}

\begin{figure*}[t!]
    \centering
    \includegraphics[width=\textwidth]{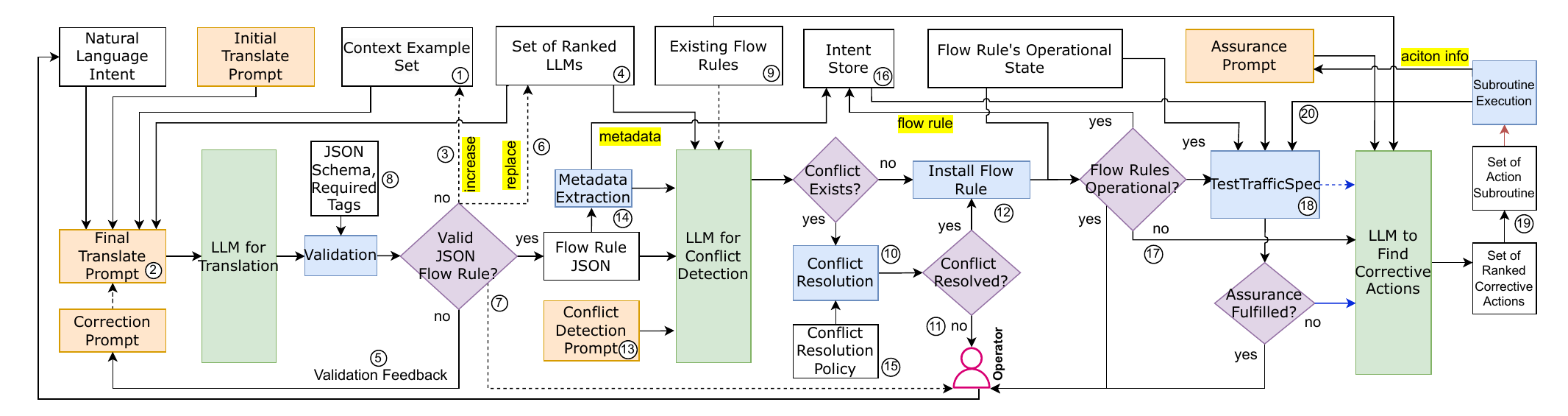}
    \caption{NetIntent architecture; end-to-end intent-based automation.}
    \label{architecture}
\end{figure*}

In Fig. \ref{architecture_abstract}, we illustrate an overview of NetIntent, the proposed end-to-end IBN system. This architecture leverages LLMs and non-LLM agents to automate the processes of intent translation, activation, and assurance. In the translation stage, user-defined natural language intents are fed into the system. 
An LLM translates the user's intent into a JSON-formatted configuration for the target SDN controller. This stage is supported by non-LLM agents, including a JSON validator that checks the syntactic and structural correctness of the LLM output, and a feedback module that prompts the LLM to refine or correct its output based on validation results. 
The next stage is intent activation which does conflict detection, conflict resolution, and intent deployment. Here, an LLM checks the newly translated configuration for conflicts against existing configurations. If a conflict is found, a resolver tries to resolve it. Once this is done, the new configuration is deployed on the target SDN controller. If this succeeds, the third stage begins, that is, intent assurance. Here, non-LLM agents provide the LLM information on intent drift and current state of the device. LLM then generates corrective actions which are executed by code modules. This is continued in a closed-loop fashion to provide intent assurance. In Fig. \ref{architecture}, we illustrate the full architecture of NetIntent. In the following sections, we describe the architecture in details.

\vspace{-0.3cm}
\subsection{NetIntent WORKFLOW and DETAILS}

\subsubsection{INTENT TRANSLATION}
NetIntent asks the network operator for intent, expressed in natural language. Supported by context examples and prompt, LLM starts translating the intent. The carefully designed prompt guides the LLM to translate intent into JSON structured configuration. The output is then fed to a validator to check for any syntax error or missing tags necessary for the target SDN controller. If the configuration is not valid, a feedback is sent back to the LLM automatically. Several data is given to the LLM to fix the error which includes the flow rule configuration, the validation feedback, and a new guiding prompt. This prompt instructs the LLM to fix the error and regenerate the corrected configuration. This process is repeated until a valid configuration is obtained. Once a valid configuration is found, it serves as the SDN-controller-readable representation of the user’s intent, prepared to be sent to the intent activation stage. In Algorithm \ref{alg:intent_translation}, we formally describe the intent translation process.

\begin{algorithm}[t!]
\caption{Intent Translation using Ranked LLMs}
\label{alg:intent_translation}
\begin{algorithmic}[1]

\Statex \textbf{Input:} User intent \(U\), context examples \(\mathcal{C}\), ranked LLMs \(\mathcal{M} = \{M_1, M_2, \dots, M_k\}\), initial context count \(x_0\), maximum context examples \(Y\)
\Statex \textbf{Output:} Valid JSON configuration \(\Phi\) or validation feedback \(\epsilon\)

\For{each LLM \(M_i \in \mathcal{M}\)}
    \State \(x \gets x_0\)
    \While{\(x \leq Y\)}
        \State Construct prompt using \(U\) and relevant-\(x\) examples from \(\mathcal{C}\)
        \State Generate output \(\Phi = M_i(\text{prompt})\)
        \State Validate \(\Phi\) against SDN controller specification
        \If{\(\Phi\) is valid}
            \State \Return \(\Phi\)
        \Else
            \State Extract validation error information \(\epsilon\)
            \State Update the correction prompt with \(\epsilon\) and \(\Phi\)
            \State \(x \gets x + 1\)
        \EndIf
    \EndWhile
\EndFor
\State \Return Report with validation feedback \(\epsilon\) to operator for manual resolution
\end{algorithmic}
\end{algorithm}

\noindent
Next, we describe the related components such as context examples, set of LLMs, and validator, which play very important roles in intent translation.

\vspace{-0.3cm}
\paragraph{CONTEXT EXAMPLE SELECTION}
\label{context_example}
The set of context examples (\circnum{1} in Fig. \ref{architecture}) is a dataset consisting of input and expected output pairs for the intent translation task. For few-shot learning of LLM, these input-output pairs are used as context examples which tell LLM how the output should look like for the corresponding input. Studies \cite{ye-etal-2023-complementary, si-etal-2023-measuring} have shown that adding more examples in the LLM prompt helps produce better output. Moreover, the more relevant the examples are to the target task, the more accurate result can be found from LLM. As outlined in Sec. V-C-1-\ref{prompt_engg}, the context examples are incorporated in the prompt (\circnum{2}). We update the number of context examples fed to the LLM prompt when necessary which is indicated by dotted line (\circnum{3}). For instance, when LLM fails to generate a valid configuration with a given set of context examples, the number of context examples is increased. In Algorithm \ref{alg:intent_translation}, we formally described it. However, there should be a trade-off between LLM output accuracy and its response time since adding more context examples to the LLM prompt increases the output generation latency. As for the selection of context examples for the target task, we use the strategy called Max Marginal Relevance example selector\cite{ye-etal-2023-complementary}. It ensures that the chosen examples are not only highly similar to the input query, but also maximally diverse from each other, preventing redundancy and improving the overall quality and coverage of the provided context. The reason we leverage dynamic context example selection for intent translation is that the input intents can be linguistically diverse and network can have thousands of different configurations. Fixed examples in prompt work well for a small number of use cases but become impractical with larger use cases due to LLM context length limitations and increased processing time. In order to provide context examples for intent translation, we use Intent2Flow-ODL and Intent2Flow-ONOS datasets.

\paragraph{LLM PROMPT FOR INTENT TRANSLATION}
\label{prompt_engg} 
Here, we describe how the prompt is designed to guide the LLM for intent translation task. A prompt is the input instruction in text form given to the LLM to elicit a desired output. The prompt we design guides the LLM to generate a JSON formatted configuration for the target SDN controller's device. We chose JSON format due to several advantages. JSON templates are widely used in configuration tools which allows for smooth integration into deployment workflows. Additionally, LLMs, trained on vast amounts of data, are already familiar with JSON examples which enhances their accuracy in generation of JSON outputs. 

The prompt we designed for intent translation consists of four parts: general instructions, the output template, examples, and user intent. The prompt begins by defining the task of converting natural language network intents into JSON-formatted flow rules for the target SDN controller. It explicitly states that the response must contain only valid JSON, with no natural language. Controller-specific nuances, such as omitting the \textit{treatment} field in ONOS to drop traffic, are clearly explained. The prompt includes a detailed JSON template that reflects the expected data model of the target controller. Mandatory and optional fields are carefully annotated. Usage rules are specified—for example, assigning high priority to queue-based flows or including VLAN fields only when explicitly required. Higher priority is also assigned for security related flows such as blocking a certain prefix. While not embedded in the prompt body, context examples are dynamically inserted at runtime after the JSON template part. The last part of the prompt contains the actual operator-provided intent. Additionally, a specialized slicing prompt is used to detect and extract slice-related metadata such as switch\_id, queue\_id, and port\_id, etc. The slicing prompt follows the same design structure. In Table \ref{tab:llm_prompt_for_translation}, the structure of the prompt is described. We provide the full prompt details on GitHub~\cite{NetIntent2025}.

\renewcommand{\arraystretch}{0.9}
\begin{table}[t!]
\caption{Structured LLM prompt input}
\label{tab:llm_prompt_for_translation}
\centering
\normalsize
\resizebox{\columnwidth}{!}{%
\begin{tabular}{p{2.5cm} >{\setstretch{0.5}}p{9.7cm}} 
\toprule
\textbf{Item} & \textbf{Description / Source} \\
\midrule
Intent & User-provided natural language description of desired network behavior or policy \\
Intent Type Classifier & Determines whether the intent relates to queuing/slicing (QoS) or not, using a rule-based or LLM-driven slicing prompt \\
Context Examples & Few-shot in-context examples drawn from the same dataset to aid translation performance \\
Controller Type & Identifier indicating the SDN controller (e.g., ODL or ONOS), used to guide schema selection \\
Schema Template & JSON policy format that matches the controller's northbound API specification (e.g., ODL-style flow entries or ONOS-style flow rules) \\
Match Fields & Derived from the intent (e.g., source/destination IPs, transport ports, IP protocol, in-port) to define traffic selectors \\
Treatment Fields & Derived from the intent (e.g., output port, drop, queue ID, VLAN tag) to define actions taken on matched traffic \\
Flow Metadata & Fields such as flow ID, priority, timeout, and flow name—either set to defaults or policy-specific values \\
Optional Fields Filter & Logic to ensure optional fields (e.g., queue, VLAN, port) are only included if explicitly mentioned in the intent \\
Validation Rules & Structural and semantic checks to ensure JSON output conforms to the controller’s schema and is conflict-free \\
Fallback Behavior & If the intent cannot be reliably translated into a policy, the LLM should return an empty JSON object \\
\bottomrule
\end{tabular}%
}
\end{table}

\vspace{-0.2cm}
\paragraph{MULTI-LLM COORDINATION}
We use more than one LLM in NetIntent (\circnum{4}) so that the repeated failure of one LLM does not put the system in an infinite loop with no progress. If there are $N$ LLMs in the system, there will be a ranking of the LLMs based on their performance. Starting from the top-ranked model, NetIntent incrementally augments the prompt with additional context examples and validation feedback (\circnum{5}) until a valid output, i.e., correct JSON formatted flow rule, is produced or the maximum context example limit is reached. In no valid output is generated and context example limit is exceeded, then the current LLM is replaced (\circnum{6}). However, if no LLM generates a valid configuration, a validation report is returned for manual intervention (\circnum{7}). In Sec. \ref{exp}, we benchmarked several LLMs which can be used to rank them. In Algorithm \ref{alg:intent_translation}, we describe how the NetIntent updates the context example and LLM dynamically.

\vspace{-0.2cm}
\paragraph{LLM OUTPUT VALIDATOR}
After an LLM generates a configuration in JSON format, it is sent through a validator to check for syntax error as well as conformity with the required tags and values for the target SDN controller. Although, LLM itself can check this, adding a separate validator further ensures that only the correct configuration is sent to the SDN controller device. NetIntent loads \circnum{8} controller-specific syntax rules and required tags, checks for structural and controller-specific syntax violations, and verifies the presence of all necessary tags. The tags needed depend on the intent type, which we describe in Sec. V-C-2-\ref{conflict_resolution}. If no errors are found, the valid configuration is returned; otherwise, a validation feedback detailing the errors and their locations is produced.

\vspace{-0.3cm}
\subsubsection{INTENT ACTIVATION}
In the intent activation stage, the generated configuration is checked for conflicts with the configurations already installed in the target SDN controller device. If the new configuration has no conflict with existing configurations, then it is installed on the target SDN controller device.

\vspace{-0.2cm}
\paragraph{CONFLICT DETECTION}
\label{section_conflict_detect}
We leverage LLM to find conflicts between pairs of JSON formatted configurations, also called flow rules. They are usually saved in the intent \textit{Configuration data store} \cite{opendaylight2015} in ODL SDN controller and in \textit{FlowRuleStore} \cite{onosprojectFlowRuleStoreONOS} in ONOS SDN controller. In Fig. \ref{architecture}, the conflict detection module retrieves existing configurations from the target SDN controller (shown in \circnum{9}) and compare against the new flow rule using LLM. For accurate conflict detection, writing an appropriate prompt is necessary. The prompt describes how to handle the JSON flow rules and what constitutes a conflict. Unlike intent translation, we use single LLM in the conflict detection system. This is based on our evaluation of LLMs on conflict detection task where we found that a well-prompted LLM can detect almost every conflict. However, if multiple LLMs are used for conflict detection, it is recommended to use them in a voting fashion where LLM's decision on conflict existence will be treated as a vote and the final decision will be based on majority votes. In Algorithm \ref{alg:conflict_detection}, we show how conflict detection is done through LLM. If any conflict if found, the system tries to resolve the conflict (\circnum{10} in Fig. \ref{architecture}) as described in Sec. V-C-2-\ref{conflict_resolution}. If it cannot be resolved, NetIntent sends the detail of the conflict to the operator (\circnum{11}) and does not attempt to install the new flow rule. In case there are no conflicts, NetIntent attempts to install the flow rule in the target SDN controller device (\circnum{12}), as described in Sec. V-C-2-\ref{flow_rule_install}.

\vspace{-0.3cm}
\paragraph{LLM PROMPT FOR CONFLICT DETECTION} 
To detect conflicts between flow rules, we designed controller-specific prompts (\circnum{13} in Fig. \ref{architecture}) that guide an LLM to perform strict, field-by-field comparisons of flow configurations in JSON format. The prompt logic is customized for each SDN controller—ODL and ONOS—based on their schema and semantics. A conflict is identified only when two conditions are simultaneously satisfied: the match criteria of both rules overlap, and their actions are contradictory. Overlap requires both flows to specify identical or overlapping values for key fields such as source and destination IP addresses, transport protocols, and port numbers. For IP fields, CIDR prefix matching is used to assess overlap; however, if any critical field is missing in either flow, the match for that field is considered non-overlapping. Once overlapping criteria are established, the prompt checks whether the actions differ in a meaningful way. For ODL, this includes mismatches in output ports, presence of a \textit{drop-action} in one rule versus an output-action in another, or conflicting queue IDs in \textit{set-queue-actions}. For ONOS, action differences are determined by inspecting the \textit{treatment} field; examples of conflict include differing output ports, presence of a queue instruction with different \textit{queueId}, or cases where one rule uses \textit{NOACTION} (interpreted as drop) while the other forwards packets. Importantly, priority is disregarded in this stage; priority determines which rule takes effect in deployment but not whether a conflict exists. Since a priority value may be set incorrectly by LLM during translation, reporting conflict based on that may produce false positives. However, we don't overlook the priority, rather we resolve priority if LLM reports a conflicts. The resolved priority takes effect in real network operation where packets are matched by the network device based on rule priority.
The LLM is instructed to return a structured JSON response with a conflict status field and an explanation field. In Table \ref{tab:llm_prompt_conflict_detection}, the structure of the prompt is described. We provide the full prompt on GitHub~\cite{NetIntent2025}.

\renewcommand{\arraystretch}{0.8}
\begin{table}[t!]
\caption{Structured LLM prompt input for conflict detection}
\label{tab:llm_prompt_conflict_detection}
\centering
\normalsize
\resizebox{\columnwidth}{!}{%
\begin{tabular}{p{3.2cm} >{\setstretch{0.5}\raggedright\arraybackslash}p{9.8cm}}
\toprule
\textbf{Item} & \textbf{Description} \\
\midrule
Flow 1 (JSON) & First flow rule in controller-specific JSON format (ODL or ONOS) \\
Flow 2 (JSON) & Second flow rule to be compared with Flow 1 \\
Match Fields & Literal comparison of fields: IP source/destination, protocol, ports, in-port, and Ethernet type \\
Wildcard Handling & Missing fields are treated as general (apply to all); mismatched fields imply no overlap \\
Conflict Conditions & Triggered only when match fields overlap \textit{and} actions contradict (e.g., drop vs. forward, different output ports or queues) \\
Exceptions & Additional match fields in one rule that are absent in the other prevent conflict declaration \\
Priority Handling & Ignored during conflict detection \\
Action Comparison & Drop vs. forward, differing queues or ports are considered conflicting if matches align \\
Conflict Output & JSON object with \texttt{conflict\_status} (0/1) and \texttt{conflict\_explanation} string \\
\bottomrule
\end{tabular}%
}
\end{table}

\vspace{-0.3cm}
\paragraph{CONFLICT RESOLUTION}
\label{conflict_resolution}
To determine which flow rule should take precedence in the event of a detected conflict, we propose a resolution policy denoted as $\mathcal{P}$. This policy defines a structured set of decision rules used to evaluate and prioritize conflicting flow configurations based on their semantic intent, specificity, and contextual metadata. This is shown in Fig. \ref{architecture} in \circnum{15}. To use $\mathcal{P}$, we first classify the flow rules. Next, we describe classification approach.

\vspace{0.1cm}
\noindent
\textbf{Flow Rule Classification:} During the conflict detection stage, each flow rule is annotated with metadata (\circnum{14} in Fig. \ref{architecture}) to support downstream reasoning and resolution. This metadata includes the rule's \textit{type} (e.g., security, forwarding, or QoS) and a numerical measure of \textit{specificity}, which reflects how narrowly scoped the rule is. To extract this metadata, we parse the flow rule translated by LLM. The metadata fields are inferred as follows:

\begin{itemize}
    \item Type: It is determined from the rule's action structure. For ODL, this includes fields within \textit{apply-actions.action} such as \textit{drop-action}, \textit{set-queue-action}, or \textit{output-action}. For ONOS, this is inferred from the \textit{treatment.instructions} field, e.g., \textit{type: NOACTION} implies \textit{security}, \textit{QUEUE} implies \textit{qos}, and \textit{OUTPUT} implies \textit{forwarding}. Also, the absence of \textit{treatment.instructions} field implies \textit{security}.
    
    \item Specificity: It is computed as the number of explicit match fields in the rule. For ODL, these are found under the \textit{match} field; for ONOS, under \textit{selector.criteria}. If the rule includes IP-based fields with CIDR notation (e.g., \textit{10.0.0.1/32}), the prefix length is normalized (e.g., $/32 \rightarrow 1.0$) and added to the specificity score to reflect its precision. 
\end{itemize}


\vspace{0.1cm}
\noindent
\textbf{Conflict Resolution Policy: } The resolution policy $\mathcal{P}$ is a deterministic rule-based mechanism designed to evaluate conflicting flow rules based on a predefined set of priorities. Using the flow rule classification data, the logic first checks whether either rule represents a security intent or not. If so, the security-related rule is favored, reflecting the importance of enforcing access control and traffic blocking over general forwarding or performance behaviors. If both rules are of the same type, the policy then compares their specificity. The more specific rule that typically targets a narrower traffic subset is preferred to ensure precise enforcement. If both rules are of equal type and specificity, the policy falls back to comparing their assigned priority values, selecting the one with the higher priority. This approach ensures that critical intents are not inadvertently overridden by broader or lower-priority rules. While this default logic aligns with common security and operational goals, it can be extended or customized to reflect domain-specific resolution strategies or business policies. In Algorithm \ref{alg:conflict_detection}, we formally describe the steps of conflict detection and resolution approach.

\begin{algorithm}[t!]
\caption{Intent Activation using LLM}
\label{alg:conflict_detection}
\begin{algorithmic}[1]
\Statex \textbf{Input:} Flow rules \(\phi_1, \phi_2\), SDN controller \(S\), LLM \(M\), resolution policy \(\mathcal{P}\)
\Statex \textbf{Output:} Conflict status, explanation, and resolution decision

\State Construct prompt for \(S\)
\State Provide \(\phi_1\), \(\phi_2\) as input to \(M\)
\State Query $M$ and parse response (\textit{conflict\_status}, \textit{conflict\_explanation})
\If{\(\textit{conflict\_status} = 1\)}
    \State Use Metadata inference algorithm to infer metadata for \(\phi_1\), \(\phi_2\)
    \State \textbf{goto Conflict Resolution}
\EndIf
\State \Return (\textit{conflict\_status} = 0,\hspace{0pt} \textit{conflict\_explanation} = ``")
\Statex \textbf{Conflict Resolution:}
\State Evaluate \(\phi_1\), \(\phi_2\) using policy \(\mathcal{P}\) with inferred metadata
\If{resolution decision cannot be made (e.g., policy \(\mathcal{P}\) yields no clear priority)}
    \State Generate detailed conflict report including \(\phi_1\), \(\phi_2\), and \textit{conflict\_explanation}
    \State Send report to network operator for manual resolution
    \State \Return (\textit{conflict\_status}=1, \hspace{0pt}\textit{conflict\_explanation}, \hspace{0pt}\textit{priority\_rule=None})

\Else
    \State Determine priority rule \(\Phi^*\) based on rule type, specificity, and priority
    \State Deploy \(\Phi^*\) to controller \(S\)
    \State Log the non-priority rule for audit or operator review
    \State Optionally notify assurance module or user of conflict resolution outcome
    \State \Return (\textit{conflict\_status} = 1,\hspace{0pt} \textit{conflict\_explanation},\hspace{0pt} \textit{priority\_rule} = $\Phi^*$)
\EndIf
\end{algorithmic}
\end{algorithm}

\vspace{-0.3cm}
\paragraph{FLOW RULE INSTALLATION}
\label{flow_rule_install}
If no conflicts are found or an identified conflict is resolved, the JSON flow rule is pushed to the target device of the target SDN controller (\circnum{12} in Fig. \ref{architecture}). For ODL, it is installed in ODL's Configuration data store via the RESTCONF API. For ONOS, it is installed in FlowRuleStore via REST API. To install the flow rule, first we infer the device ID of the target SDN controller where the rule will be installed from the input intent. Then we use POST method for ONOS and PUT method for ODL to install the flow rule to the target device. If the installation succeeds without any error, we verify if the rule is reflected on the target device. For ODL, we check if it is present in ODL's Operational data store while for ONOS, we check if it is present in FlowRuleStore. We keep a copy of the installed intent, its JSON flow rule, metadata, along with the target device information in our system. We save it in a file called ``IntentStore" for future reference. The IntentStore is updated with the insertion and deletion of flow rules in the system.

\vspace{-0.6cm}
\subsubsection{INTENT ASSURANCE}
Intent assurance is the mechanism by which an IBN system continuously validates and enforces that the network state and behavior align with the user-defined intents. To implement it, the intent assurance module of NetIntent does the following tasks: 1) Verify that the flow rules generated from user intents are correctly applied on all relevant devices by utilizing real-time traffic. 2) Trigger corrective actions if any intent is not being fulfilled. 

These tasks are repeated in a closed-loop fashion for continuous monitoring of the system. Below, we describe them in details. In Algorithm \ref{alg:closed_loop_assurance}, we formally describe the closed-loop assurance module of NetIntent. Next, we describe each step in details.

\begin{algorithm}[htbp]
\caption{Closed-Loop Intent Assurance using LLM}
\label{alg:closed_loop_assurance}
\begin{algorithmic}[1]
\Statex \textbf{Input:} Intent, IntentMetadata, NodeID, TableID, FlowID, [QueueID], TestTrafficSpec, ControllerInfo, MaxAttempts
\Statex \textbf{Output:} Verified Intent or User Alert

\For{attempt = 1 to MaxAttempts}
    \State Retrieve \textit{FlowRuleJSON}, \textit{IntentType} from \textit{IntentStore}
    \State {\raggedright Retrieve installed flow rule using \textit{NodeID}, \textit{TableID}, \textit{FlowID}\par}

    \If {rule missing or mismatched}
        \State Reinstall rule from \textit{IntentStore}
        \State \textbf{continue}
    \EndIf

    \State Retrieve $S_{\text{initial}}$ (packet, byte, and queue stats if applicable)
    \State Transmit test traffic; record $T_{\text{start}}$, $T_{\text{end}}$
    \State Retrieve $S_{\text{final}}$ and compute $\Delta S = S_{\text{final}} - S_{\text{initial}}$

    \If {\textit{IntentType} = ``Forwarding"}
        \If {$\Delta S.\textit{packet-count}<$\textit{ExpectedPacketCount}}
            \State \textbf{goto} LLM-Remediation
        \Else
            \State Log: ``Forwarding verified''; \Return Verified
        \EndIf

    \ElsIf {\textit{IntentType} = ``Security"}
        \If {rule has forwarding behavior (ODL: \textit{output-action}, ONOS: not \textit{NOACTION})}
            \State \textbf{goto} LLM-Remediation
        \ElsIf {$\Delta S.\textit{packet-count}  <$\textit{ExpectedBlockedPackets}}
            \State \textbf{goto} LLM-Remediation
        \Else
            \State Log: ``Security verified''; \Return Verified
        \EndIf

    \ElsIf {\textit{IntentType} = ``QoS"}
        \State Let $B_t$, $R_t$ = expected byte count, rate; $\Delta B$ = change in \textit{tx\_bytes}[\textit{QueueID}]
        \State $R_{\text{measured}} = (\Delta B \times 8) / (T_{\text{end}} - T_{\text{start}})$
        \If {$(\Delta B < \alpha B_t)$ or $(|R_{\text{measured}} - R_t| > \epsilon)$}
            \State \textbf{goto} LLM-Remediation
        \Else
            \State Log: ``QoS verified''; \Return Verified
        \EndIf
    \EndIf

    \State \textbf{LLM-Remediation:}
    \State Compile current context: FlowRule, DeviationMetrics, TestTrafficSpec, ControllerInfo
    \State Query LLM with context for ranked root causes and recommended actions
    \For{each action $a$ in ranked action list}
        \State Map $a$ to corresponding subroutine
        \State Execute subroutine
        \State \textbf{break} (to retry assurance in next iteration)
    \EndFor
\EndFor
\State Escalate to operator.
\end{algorithmic}
\end{algorithm}

\vspace{-0.2cm}
\paragraph{TEST TRAFFIC GENERATION}
\label{test_traffic_generation}
The \textit{TestTrafficSpec} module is responsible for generating synthetic traffic tailored to the intent under verification. In Fig. \ref{architecture}, it is shown near \circnum{18}. It provides a structured interface for specifying the traffic profile required to exercise an installed flow rule and serves as the foundation for capturing expected packet and byte-level behavior in the data plane. Operating in coordination with the assurance engine, the module produces traffic that aligns with the flow rule’s match criteria, directionality, and semantic purpose (e.g., forwarding, blocking, or QoS enforcement).

\noindent
\textbf{Forwarding Intents:} For intents involving basic packet forwarding, simple ICMP \textit{ping} is sufficient. The number of packets transmitted is explicitly specified in the \textit{TestTrafficSpec}, enabling direct comparison with the observed \textit{packet-count} delta ($\Delta S.\textit{packet-count}$) for the flow rule.

\noindent
\textbf{Security (drop) Intents:} These are verified similarly using ICMP \textit{ping} to ensure that packets match the drop rule and are not forwarded. This lightweight approach provides deterministic control over packet generation with minimal overhead.

\noindent
\textbf{QoS (queue) Intents:} For intents involving queue assignment or traffic prioritization, \textit{ping} is insufficient. The module uses TCP-based synthetic traffic (e.g., generated via tools like \textit{iperf}) to verify both byte volume and data rate. The traffic is generated with a predefined size ($B_t$) and duration, and the observed throughput ($R_{\text{measured}}$) is compared to the expected rate ($R_t$) as part of assurance.

To support controller-agnostic deployment, the \textit{TestTrafficSpec} includes source/destination resolution, traffic characteristics (e.g., volume, duration, protocol), and expected performance metrics.
\vspace{-0.3cm}
\paragraph{IDENTIFYING INTENT DRIFT}
NetIntent identifies intent drift by dealing with the dual challenge of verifying that (i) high-level intents have been correctly translated and deployed as low-level flow rules, and (ii) these flow rules are actively enforcing the intended behavior in the data plane.

These challenges becomes more nuanced across diverse intent types, such as forwarding, security, and QoS, each of which requires distinct criteria and metrics for effective verification. To address this, we maintain a storage called \textit{IntentStore} that retains the original natural language intent, its translated JSON flow rule, and relevant metadata (e.g., intent type, specificity, and expected traffic characteristics). It is shown in Fig. \ref{architecture} \circnum{16}. During the assurance stage, we iterate over each intent and retrieve the corresponding flow rule using its unique identifiers: \textit{NodeID}, \textit{TableID}, and \textit{FlowID} from the target SDN controller. We verify that the deployed rule is present and structurally consistent with its stored specification. If not, a corrective action is sought immediately using LLM (\circnum{17}). In Sec. V-C-2-\ref{trigger_correction}, we describe how corrective actions is taken.

Upon confirming rule presence, we collect its initial operational statistics, including \textit{packet-count} (or \textit{packets}), \textit{byte-count} (or \textit{bytes}), and where applicable, queue-level counters retrieved from the data plane via standard switch-level interfaces (e.g., Open Virtual Switch (OVS)). We denote the statistics collected prior to traffic generation as $S_{\text{initial}}$ and those collected afterward as $S_{\text{final}}$. Their difference, $\Delta S = S_{\text{final}} - S_{\text{initial}}$, reflects the observed activity of the rule over the assurance window.

To exercise the rule, we generate synthetic test traffic via our \textit{TestTrafficSpec} module, as described in Sec. V-C-3-\ref{test_traffic_generation}, which produces packets matched to the flow rule’s criteria and includes an expected byte volume $B_t$, expected packet count $P_t$, and expected transmission rate $R_t$. After the traffic completes, we analyze $\Delta S$ in conjunction with $B_t$, $P_t$, and $R_t$ to determine whether the intent has been correctly enforced. The evaluation criteria depend on the type of intent under test, as described next.

\noindent
\textbf{Forwarding Intents:} For forwarding intents, we validate whether the installed flow rule actively forwards matching packets. Let $S_{\text{initial}}$ and $S_{\text{final}}$ denote the flow's operational statistics before and after test traffic, respectively. We define the packet-count delta as $\Delta S.\textit{packet-count} = S_{\text{final}}.\textit{packet-count} - S_{\text{initial}}.\textit{packet-count}$. We expect this to align with the expected test packet count, denoted $P_t = \textit{ExpectedPacketCount}$, derived from the test specification. A value of $\Delta S.\textit{packet-count} < P_t$ may indicate forwarding misbehavior due to rule shadowing, incorrect match fields, or silent drops.

\vspace{0.2em}
\noindent
\textbf{Security Intents:} For drop or block intents, assurance requires confirming that packets match the rule but are not forwarded. We verify that $\Delta S.\textit{packet-count}$ increases, indicating that traffic was matched. Drop behavior is determined by rule semantics: in ODL, a drop rule explicitly includes \textit{drop-action:\{\}} in the \textit{apply-actions} part; in ONOS, it is indicated by the presence of only \textit{NOACTION} in the \textit{treatment.instructions} list. A rule is considered verified for a security intent if it satisfies these controller-specific drop semantics and $\Delta S.\textit{packet-count} \geq B_p$, where $B_p = \textit{ExpectedBlockedPackets}$ is the expected number of dropped packets.

\vspace{0.2em}
\noindent
\textbf{QoS Intents:} To validate QoS intents, we ensure that traffic matching the flow rule is correctly classified into the assigned queue and forwarded with expected volume and rate. Let $B_{\text{before}}$ and $B_{\text{after}}$ be the queue's \textit{tx\_bytes} counter before and after test traffic. The byte-count delta is defined as $\Delta B = B_{\text{after}} - B_{\text{before}}$. The test specification defines $B_t = \textit{ExpectedByteCount}$ and $R_t = \textit{ExpectedRate}$ as the expected volume and rate. We record the start and end timestamps $T_{\text{start}}$ and $T_{\text{end}}$ to calculate the observed data rate as $R_{\text{measured}} = (\Delta B \times 8) / (T_{\text{end}} - T_{\text{start}})$. A QoS intent is considered successfully enforced if both the volume and rate satisfy:$\Delta B \geq \alpha \times B_t \quad \text{and} \quad |R_{\text{measured}} - R_t| \leq \epsilon$, where $\alpha$ is a volume margin factor (e.g., 0.98) and $\epsilon$ is a tolerance for rate deviation.

\vspace{0.2cm}
\noindent
This validation framework supports both ODL and ONOS while it can be extended to other controllers. It enables dynamic intent assurance across SDN platforms. In Algorithm \ref{alg:closed_loop_assurance}, we formally describe the intent assurance stage of NetIntent. This assurance mechanism enables the system to automatically identify stale, misbehaving, or misconfigured intents by correlating observed network behavior with expected outcomes defined in the test traffic specification. Upon detection of inconsistency between control plane configuration and data plane behavior, corrective actions are triggered to restore alignment. 
\vspace{-0.3cm}
\paragraph{TRIGGERING CORRECTIVE ACTIONS}
\label{trigger_correction}

If an intended rule does not exists in the target device, then the rule will be reinstalled using the information of \textit{IntentStore}. However, if the rule exists and found to be identical to the originally installed one but it fails to meet its expected behavior during assurance, it becomes necessary to identify the root cause and apply corrective actions. While deterministic mappings from intent types to static fixes can handle known failure modes, they fall short when facing ambiguous behaviors or misconfigurations in underlying infrastructure such as queue bindings or port mappings. To overcome this, we integrate LLM as an intelligent diagnostic component that can reason over structured context and recommend ranked corrective actions.

\noindent
\textbf{Prompting Strategy: } To enable contextual reasoning, the LLM is provided with a prompt that includes relevant data as shown in Table \ref{tab:llm_prompt_for_corrective_act}. The LLM is instructed to:

\renewcommand{\arraystretch}{0.8}
\begin{table}[t!]
\caption{Structured LLM prompt input for generating corrective actions}
\label{tab:llm_prompt_for_corrective_act}
\centering
\normalsize
\resizebox{\columnwidth}{!}{%
\begin{tabular}{p{3.2cm} >{\setstretch{0.5}\raggedright\arraybackslash}p{9.8cm}}
\toprule
\textbf{Item} & \textbf{Description} \\
\midrule
Intent & Natural language description and metadata from \textit{IntentStore} \\
Flow Rule & JSON-encoded rule deployed for the intent \\
TestTrafficSpec & Expected traffic behavior (packet count, byte volume, rate, source/destination) \\
Deviation & Summary of assurance failure metrics from $\Delta S$ \\
Installed Rules & JSON list of all flow rules installed on the device (from controller API) \\
Queue Stats & Switch-level QoS configuration (via \textit{ovs-vsctl list qos/queue}) \\
Controller & Controller type identifier (e.g., ODL or ONOS) \\
Optional Feedback & Previous corrective actions and whether they succeeded \\
\bottomrule
\end{tabular}%
}
\end{table}

\begin{itemize}
    \item Analyze the given context to identify potential root causes of the observed deviation.
    \item Rank the causes by likelihood and severity.
    \item Propose corrective actions for each cause.
    \item Output in structured format (e.g., JSON or numbered list) to support automated parsing.
\end{itemize}

\noindent
\textbf{Corrective Action Execution Engine: } The LLM's output is parsed and mapped to a library of predefined actionable routines (\circnum{19} on Fig. \ref{architecture}). These routines correspond to common remediation procedures mentioned in Table \ref{tab:trigger_action}.

\renewcommand{\arraystretch}{0.8}
\begin{table}[t!]
\caption{Common action types and descriptions}
\label{tab:trigger_action}
\centering
\normalsize
\resizebox{\columnwidth}{!}{%
\begin{tabular}{p{3.2cm} >{\setstretch{0.5}\raggedright\arraybackslash}p{9.8cm}}
\toprule
\textbf{Action Type} & \textbf{Description} \\
\midrule
check\_match\_fields & Re-evaluate the accuracy of source/destination, ports, and protocol in the flow rule \\
increase\_priority & Raise flow rule priority to avoid shadowing \\
verify\_queue\_mapping & Check if the flow’s \textit{queue-id} is valid and mapped to the correct port \\
retranslate\_intent & Use the LLM to regenerate a refined version of the flow rule \\
remove\_output\_action & For security intents, remove unintended forwarding behavior \\
\bottomrule
\end{tabular}%
}
\end{table}

The actions are executed in the order ranked by the LLM. After each action (or set of actions), the assurance process is rerun to measure whether the deviation is resolved (\circnum{20}). If not, the failed result is appended to the prompt for the next round, enabling a feedback loop where the LLM can refine its hypothesis and suggest deeper diagnostics.

\noindent
\textbf{Closing Loop: } The corrective framework described above is realized as a closed-loop control process, summarized in Algorithm~\ref{alg:closed_loop_assurance}. The loop continues and the network operator is informed of the intent assurance updates. This ensures adaptive and context-aware remediation based on real-time deviation analysis and feedback from prior attempts.
\vspace{-0.6cm}
\subsubsection{NetIntent's EXTENSIBILITY TO OTHER SDN CONTROLLERS}
NetIntent is designed with modularity and controller-agnostic principles, making it extensible to a wide range of SDN controllers beyond ODL and ONOS. Its architecture separates controller-specific logic from core functionalities like intent translation, conflict detection, and assurance. To adapt NetIntent to another SDN controller, one would need to develop a controller-specific schema template that defines the required JSON structure for flow rules, compatible with the new controller's northbound API. Additionally, the validator component should be updated to incorporate the syntax rules and required tags specific to the new controller. By leveraging the existing modular components and updating only the controller-specific modules, NetIntent can be effectively extended to support other SDN environments.

\vspace{-0.3cm}
\section{EXPERIMENTAL RESULTS AND ANALYSIS}
\label{exp}

\subsection{HARDWARE SETUP}
To evaluate the open-source LLMs, we utilized an AMD Ryzen Threadripper PRO 5995WX 64-core processor, 500 GB of RAM, and an NVIDIA RTX A6000 GPU with 48 GB of VRAM. 

\vspace{-0.3cm}
\subsection{SOFTWARE USED FOR IMPLEMENTATION}
\label{topology_setup}

\begin{figure}[t!]
    \includegraphics[width=.9\linewidth]{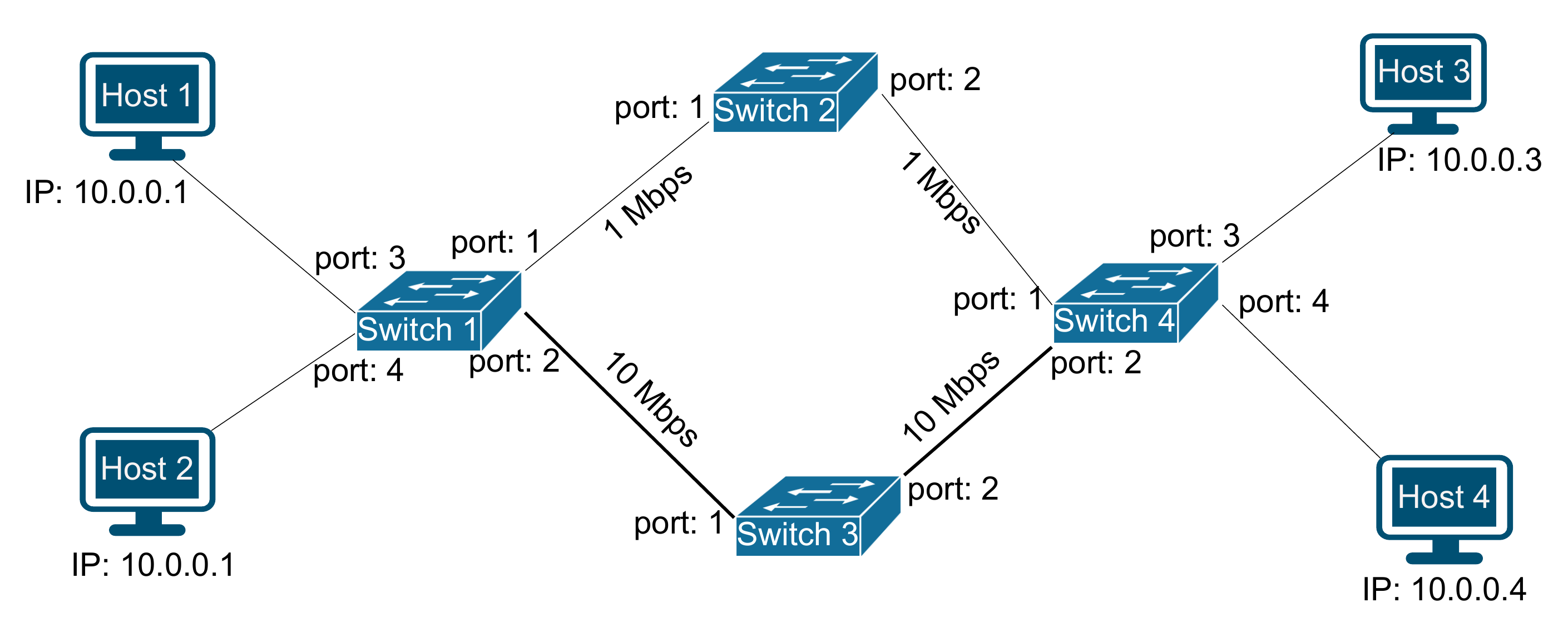}
    \caption{Topology used for flow rule installation}
    \label{topology}
\end{figure}

We implemented NetIntent using Python programming language. The LangChain \cite{topsakal2023creating} library was used to manage all LLM operations. For ODL installation, Karaf-0.8.4 was used and for ONOS, version 2.0.0 was used. We used Mininet \cite{dholakiya2021survey} to define a diamond-shaped network topology shown in Fig. \ref{topology}. The topology comprises four switches (\texttt{s1}–\texttt{s4}) and four hosts (\texttt{h1}–\texttt{h4}), where \texttt{h1} and \texttt{h2} are connected to \texttt{s1}, and \texttt{h3} and \texttt{h4} are connected to \texttt{s4}. The topology provides two distinct paths between \texttt{s1} and \texttt{s4}: one via \texttt{s2} with lower bandwidth links (1 Mbps), and another via \texttt{s3} with higher bandwidth links (10 Mbps). To simulate QoS behavior, we created traffic queues on \texttt{s3} using Mininet, assigning 6 Mbps (Queue 0) and 4 Mbps (Queue 1) to two different forwarding paths. This topology was implemented separately for both ODL and ONOS SDN controllers, and used throughout our experiments to evaluate intent translation, conflict detection, QoS policy enforcement and assurance.

\vspace{-0.3cm}
\subsection{LLM HYPERPARAMETER}
The LLMs were downloaded from OLlama~\cite{oLlama} for experiments. For intent translation tasks, the temperature of LLM was set to 0.6 and top-p was set to 0.3, while for conflict detection tasks, the temperature was set to 0.3 and top-p to 0.5. These settings were chosen to balance accuracy and variability, but they remain flexible and adjustable for future experimentation or deployment needs.

\vspace{-0.3cm}
\subsection{LLM BENCHMARKING RESULTS FOR INTENT TRANSLATION}
Here we present the result of natural language intent translation using the LLMs listed in Table \ref{tab:llm_model_list}. First we report the accuracy of translation for a dataset, then we sort out the best performing LLMs based on accuracy. 

\noindent
\textbf{Formal Specification Dataset: } We start with the Formal specification dataset. We mentioned earlier that this is a dataset of natural language intent and JSON formatted translation pair for intent covering three different requirements: reachability, waypoints and loadbalancing. To check accuracy of translation we used the verification tool provided by \cite{tu2025intent} to validate that the LLMs output JSON structure conforms to the testset data structure. Here the expected JSON is compared against the LLM-generated output JSON. The comparison is performed field by field and value by value. Each mismatch or absence of a field or value in the output is considered one mistake. The overall accuracy is calculated across all test cases, taking into account the total number of expected fields and values. Table \ref{tab:Translation_formal} shows the accuracy and average run time of 32 LLM. We did the test for all 33 LLMs but the LLM Deepseek-coder-v2:16b failed to produce meaningful result as it does not support ``K-shift" which is required for precessing long prompts.

\begin{tcolorbox}[colback=gray!10!white, colframe=gray!80!black, 
  boxrule=0.5pt, arc=4pt, auto outer arc,
  left=6pt, right=6pt, top=6pt, bottom=6pt]
\textbf{Highlights:}
Large models like QwQ-fusion and Command-r achieve high accuracy ($\geq 99\%$) on the Formal specification dataset, while mid-sized models such as Codellama:7b and Mistral:7b perform strongly (up to 95\%) with lower latency, especially when supported by in-context examples. The dataset’s structured, schema-driven format and tightly scoped prompt favor models trained for code or structured output; smaller models often fail to maintain JSON correctness or field alignment.
\end{tcolorbox}

\renewcommand{\arraystretch}{0.6}  
\begin{table*}[t!]
\caption{Benchmarking of LLMs for intent translation for Formal specification dataset (\textbf{Ctx n} denotes n context examples)}
\label{tab:Translation_formal}
\centering
\tiny
\resizebox{\textwidth}{!}{%
\begin{tabular}{ll|cc|cc|cc|cc|cc}
\toprule
\textbf{Sl.} & \textbf{LLM} & \multicolumn{2}{c|}{\textbf{Ctx 0}} & \multicolumn{2}{c|}{\textbf{Ctx 1}} & \multicolumn{2}{c|}{\textbf{Ctx 3}} & \multicolumn{2}{c|}{\textbf{Ctx 6}} & \multicolumn{2}{c}{\textbf{Ctx 9}} \\
\cmidrule(lr){3-4} \cmidrule(lr){5-6} \cmidrule(lr){7-8} \cmidrule(lr){9-10} \cmidrule(l){11-12}
& & Acc. & Time & Acc. & Time & Acc. & Time & Acc. & Time & Acc. & Time \\
\midrule
1 & \texttt{Codegemma:7b} & 71.00 & 1.60 & 75.33 & 2.20 & 78.70 & 2.50 & 76.29 & 3.50 & 68.91 & 4.30 \\
2 & \texttt{Codellama:7b} & 75.22 & 1.40 & 81.27 & 1.30 & 83.63 & 1.50 & 82.34 & 2 & 81.57 & 2.30 \\
3 & \texttt{Codellama:34b} & 89.80 & 5.60 & 88.21 & 5.30 & 91.72 & 5.80 & 91.87 & 7.20 & 92.69 & 8 \\
4 & \texttt{Codestral:22b} & 87.14 & 3.7 & 98.52 & 4.3 & 99.23 & 4 & 98.81 & 5.1 & 95.02 & 5.9 \\
5 & \texttt{Command-r:35b} & 96.84 & 4.7 & 96.97 & 4.2 & 98.13 & 5.3 & 97.52 & 6.5 & 97.17 & 7.4 \\
6 & \texttt{Deepseek-coder:1.3b} & 1.67 & 0.4 & 26.43 & 0.5 & 39.36 & 0.6 & 40.89 & 0.8 & 42.5 & 1 \\
7 & \texttt{Dolphin-Mistral:7b} & 83.93 & 1.2 & 92.64 & 1.3 & 93.51 & 1.5 & 92.24 & 1.9 & 91.89 & 2.2 \\
8 & \texttt{Gemma2:27b} & 98.01 & 4.9 & 97.42 & 4.8 & 98.28 & 4.4 & 98.33 & 5.3 & 98.23 & 6.1 \\
9 & \texttt{Llama2:7b} & 24.33 & 3.6 & 23.46 & 1.9 & 38.63 & 1.9 & 41.91 & 2.1 & 53.92 & 3 \\
10 & \texttt{Llama3:8b} & 54.2 & 1.2 & 88.96 & 1.7 & 90.4 & 1.5 & 94.13 & 1.8 & 94.88 & 2 \\
11 & \texttt{Llama3.1:8b} & 56.59 & 1.4 & 90.57 & 1.8 & 95.65 & 1.6 & 95.78 & 1.9 & 95.47 & 2.1 \\
12 & \texttt{Llama3.2:3b} & 84.29 & 1 & 88.41 & 0.9 & 90.43 & 0.9 & 88.81 & 1.1 & 89.2 & 1.2 \\
13 & \texttt{Llava-Llama3:8b} & 22.47 & 1.3 & 66.46 & 1.4 & 78.22 & 1.5 & 82.62 & 1.9 & 81.65 & 2.1 \\
14 & \texttt{Marco-o1:7b} & 91.66 & 1.3 & 95.27 & 1.5 & 96.87 & 1.7 & 96.87 & 2 & 96.72 & 2.2 \\
15 & \texttt{Mistral:7b} & 80.05 & 1.3 & 93.61 & 1.4 & 94.05 & 1.6 & 93.78 & 2 & 94.15 & 2.3 \\
16 & \texttt{Mistral-nemo:12b} & 91.06 & 1.8 & 93.89 & 2.5 & 95.99 & 2.3 & 95.6 & 2.8 & 94.55 & 3.2 \\
17 & \texttt{Openchat:7b} & 71.09 & 1 & 91.82 & 1.4 & 93.68 & 1.6 & 93.53 & 1.9 & 93.08 & 2.2 \\
18 & \texttt{Orca-mini:3b} & 21.73 & 1.1 & 47.32 & 0.8 & 56.83 & 1.1 & 52.68 & 1.7 & 41.89 & 1.5 \\
19 & \texttt{Phi:7b} & 34.92 & 0.8 & 50.96 & 0.9 & 54.2 & 0.9 & 60.36 & 0.9 & 36.68 & 1.3 \\
20 & \texttt{Phi3:8b} & 65.42 & 1.1 & 67.78 & 1 & 70.88 & 1.1 & 20.11 & 1.2 & 35.38 & 1.3 \\
21 & \texttt{Qwen:4b} & 0.86 & 6.7 & 21.79 & 1.5 & 22.29 & 2.3 & 10.92 & 3.4 & 7.44 & 3.4 \\
22 & \texttt{Qwen2:7b} & 87.43 & 1.2 & 93.33 & 1.3 & 94.5 & 1.5 & 94.83 & 1.8 & 91.24 & 2 \\
23 & \texttt{Qwen2.5:7b} & 88.46 & 1.2 & 94.78 & 1.4 & 95.97 & 1.6 & 95.97 & 1.9 & 96.94 & 2.2 \\
24 & \texttt{QwQ:32b} & 99.43 & 5.4 & 99.25 & 4.8 & 99.38 & 5.4 & 98.03 & 6.5 & 98.76 & 7.5 \\
25 & \texttt{QwQ-abliterated:32b} & 99.65 & 5.4 & 98.98 & 4.8 & 99 & 5.5 & 98.48 & 6.5 & 99.05 & 7.5 \\
26 & \texttt{QwQ-fusion:32b} & 99.58 & 4.9 & 98.42 & 4.8 & 99.43 & 5.5 & 98.93 & 6.6 & 99.13 & 7.5 \\
27 & \texttt{Starcoder:3b} & 0 & 1.3 & 0 & 1.3 & 0 & 1.2 & 0 & 1.3 & 0 & 1.3 \\
28 & \texttt{Starcoder2:3b} & 0 & 1 & 0 & 1.1 & 0 & 1 & 0 & 1.2 & 0 & 1 \\
29 & \texttt{TinyLlama:1.1b} & 3.11 & 0.8 & 28.24 & 0.6 & 35.8 & 0.7 & 42.88 & 1.1 & 41.6 & 1.7 \\
30 & \texttt{Wizardlm2:7b} & 88.55 & 1.5 & 92.91 & 1.5 & 94.3 & 1.7 & 93.99 & 2 & 92.86 & 2.3 \\
31 & \texttt{Yi:6b} & 55.8 & 1.7 & 83.4 & 1.6 & 86.89 & 1.7 & 79.91 & 2 & 58.88 & 2.3 \\
32 & \texttt{Zephyr:7b} & 52.11 & 1.2 & 85.07 & 1.4 & 87.59 & 1.8 & 89.76 & 2.2 & 90.08 & 2.5 \\
\bottomrule
\end{tabular}%
}
\end{table*}

Table \ref{tab:Translation_formal} showcases trends in runtime, accuracy, and the impact of context examples. Larger models, such as QwQ (32b), QwQ-abliterated (32b), and Command-r (35b), consistently achieve the highest accuracy, exceeding 99\% in some cases, due to their extensive parameter capacity, enabling them to better generalize and handle complex patterns in the dataset. However, these models also demonstrate higher runtime, highlighting a trade-off between accuracy and computational efficiency. Smaller models, like TinyLlama (1.1b) and Deepseek-coder (1.3b), show significantly lower accuracy, especially with fewer context examples, due to limited capacity for nuanced understanding. The inclusion of context examples plays a pivotal role in improving accuracy across all models, with gains most pronounced in mid-sized models such as Codellama:7b and Mistral:7b, where accuracy increases by up to 20\% from zero to nine context examples. This indicates that semantically relevant and diverse examples help LLMs refine predictions and minimize errors for intent translation task. Interestingly, some models, like Llama3.1 (8b), exhibit a sharp improvement with context examples, suggesting better optimization for in-context learning. However, the diminishing returns in accuracy observed for larger models like QwQ at higher context sizes may reflect saturation in learning capacity or increased task complexity due to larger input contexts.
This result underscores the importance of balancing model size, runtime, and the careful selection of context examples to optimize both accuracy and efficiency in practical scenarios. Larger models excel in complex tasks, but the improvement in smaller models with in-context learning highlights the potential for tailored optimizations in resource-constrained settings. In Table \ref{Translation_formal_best} we present the best LLMs in terms of accuracy for Formal specification Dataset with respect to different context examples.

\renewcommand{\arraystretch}{0.6}
\begin{table}[t!]
\caption{Best LLMs for intent translation for Formal specification dataset}
\label{Translation_formal_best}
\centering
\small
\resizebox{\columnwidth}{!}{%
\begin{tabular}{clcc}
\toprule
\textbf{Context} & \textbf{LLM} & \textbf{Accuracy (\%)} & \textbf{Avg. Time (s)} \\
\midrule
\multirow{3}{*}{0} & \texttt{QwQ-abliterated:32b} & 99.65 & 5.4 \\
                   & \texttt{QwQ-fusion:32b}      & 99.58 & 4.9 \\
                   & \texttt{QwQ:32b}             & 99.43 & 5.4 \\
\midrule
\multirow{3}{*}{1} & \texttt{QwQ:32b}             & 99.25 & 4.8 \\
                   & \texttt{QwQ-abliterated:32b} & 98.98 & 4.8 \\
                   & \texttt{Codestral:22b}       & 98.52 & 4.3 \\
\midrule
\multirow{3}{*}{3} & \texttt{QwQ-fusion:32b}      & 99.43 & 5.5 \\
                   & \texttt{QwQ:32b}             & 99.38 & 5.4 \\
                   & \texttt{Codestral:22b}       & 99.23 & 4.0 \\
\midrule
\multirow{3}{*}{6} & \texttt{QwQ-fusion:32b}      & 98.93 & 6.6 \\
                   & \texttt{Codestral:22b}       & 98.81 & 5.1 \\
                   & \texttt{QwQ-abliterated:32b} & 98.48 & 6.5 \\
\midrule
\multirow{3}{*}{9} & \texttt{QwQ-fusion:32b}      & 99.13 & 7.5 \\
                   & \texttt{QwQ-abliterated:32b} & 99.05 & 7.5 \\
                   & \texttt{QwQ:32b}             & 98.76 & 7.5 \\
\bottomrule
\end{tabular}%
}
\end{table}

\noindent
\textbf{NFV Configuration Dataset: } Now we report the result of running the LLMs on  NFV configuration dataset. 

\begin{tcolorbox}[colback=gray!10!white, colframe=gray!80!black, 
  boxrule=0.5pt, arc=4pt, auto outer arc,
  left=6pt, right=6pt, top=6pt, bottom=6pt]
\textbf{Highlights:} 
In the NFV configuration translation task, context examples notably boost accuracy, and several smaller and mid-sized models (e.g., Qwen2:7b, Codellama:7b) reach very high accuracy with low latency. This shows that the rigid, narrowly scoped NFV intents allow efficient models to compete, underscoring the roles of task simplicity, prompt design, and in-context learning.
\end{tcolorbox}

The dataset has 120 natural language intent and JSON structured translation pairs. As before, half of it was chosen for providing context examples and other half was used for testing. Table \ref{tab:translation_NFV} shows the accuracy and average run time of 33 LLMs. The accuracy was determined the same as was used for formal specification dataset. The table indicate that larger models, such as Command-r:35b and Codellama:34b, demonstrate high accuracy (up to 100\%) with increased context examples, leveraging their greater parameter capacity to handle the complexity of NFV configurations. However, their higher accuracy comes at the cost of increased runtime, reflecting the trade-off between model sophistication and computational efficiency. Conversely, smaller models like TinyLlama:1.1b and Orca-mini:3b struggle with accuracy across all contexts, indicating limited generalization capabilities due to their lower parameter sizes. 

\renewcommand{\arraystretch}{0.6}
\begin{table*}[t!]
\caption{Benchmarking of LLMs for intent translation for NFV configuration dataset (\textbf{Ctx n} denotes n context examples)}
\label{tab:translation_NFV}
\centering
\tiny
\resizebox{\textwidth}{!}{%
\begin{tabular}{ll|cc|cc|cc|cc|cc}
\toprule
\textbf{Sl.} & \textbf{LLM} & \multicolumn{2}{c|}{\textbf{Ctx 0}} & \multicolumn{2}{c|}{\textbf{Ctx 1}} & \multicolumn{2}{c|}{\textbf{Ctx 3}} & \multicolumn{2}{c|}{\textbf{Ctx 6}} & \multicolumn{2}{c}{\textbf{Ctx 9}} \\
\cmidrule(lr){3-4} \cmidrule(lr){5-6} \cmidrule(lr){7-8} \cmidrule(lr){9-10} \cmidrule(l){11-12}
& & Acc. & Time & Acc. & Time & Acc. & Time & Acc. & Time & Acc. & Time \\
\midrule
1 & \texttt{Codegemma:7b} & 75.00 & 1.30 & 77.50 & 1.20 & 75.00 & 1.10 & 90.00 & 1.20 & 90.00 & 1.20 \\
2 & \texttt{Codellama:34b} & 17.50 & 3.40 & 82.50 & 3.30 & 85.00 & 3.50 & 82.50 & 4.00 & 87.50 & 4.20 \\
3 & \texttt{Codellama:7b} & 45.00 & 1.00 & 67.50 & 1.00 & 70.00 & 1.00 & 80.00 & 1.00 & 95.00 & 1.10 \\
4 & \texttt{Codestral:22b} & 72.50 & 2.80 & 72.50 & 2.20 & 82.50 & 2.30 & 87.50 & 2.60 & 95.00 & 2.90 \\
5 & \texttt{Command-r:35b} & 65.00 & 3.20 & 90.00 & 3.20 & 87.50 & 3.50 & 97.50 & 3.70 & 100.00 & 4.00 \\
6 & \texttt{Deepseek-coder-v2:16b} & 60.00 & 1.30 & 85.00 & 1.00 & 80.00 & 1.10 & 92.50 & 1.30 & 95.00 & 1.40 \\
7 & \texttt{Deepseek-coder:1.3b} & 0.00 & 1.00 & 13.51 & 0.60 & 47.06 & 0.60 & 37.14 & 0.60 & 36.84 & 0.50 \\
8 & \texttt{Dolphin-Mistral:7b} & 40.00 & 1.30 & 75.00 & 1.00 & 62.50 & 1.00 & 92.50 & 1.00 & 92.50 & 1.10 \\
9 & \texttt{Gemma2:27b} & 65.00 & 3.20 & 72.50 & 2.30 & 75.00 & 2.20 & 82.50 & 2.60 & 92.50 & 2.90 \\
10 & \texttt{Llama2:7b} & 5.00 & 2.20 & 62.50 & 0.90 & 58.97 & 1.00 & 63.89 & 1.10 & 60.00 & 1.40 \\
11 & \texttt{Llama3:8b} & 57.50 & 1.00 & 75.00 & 0.90 & 85.00 & 0.90 & 95.00 & 1.00 & 97.50 & 1.10 \\
12 & \texttt{Llama3.1:8b} & 47.50 & 1.10 & 77.50 & 1.00 & 75.00 & 0.90 & 92.50 & 1.10 & 97.50 & 1.20 \\
13 & \texttt{Llama3.2:3b} & 15.00 & 1.50 & 50.00 & 0.70 & 70.00 & 0.70 & 72.50 & 0.60 & 75.00 & 0.70 \\
14 & \texttt{Llava-Llama3:8b} & 0.00 & 1.60 & 38.46 & 1.00 & 52.78 & 0.90 & 51.28 & 1.10 & 47.37 & 1.20 \\
15 & \texttt{Marco-o1:7b} & 12.50 & 1.00 & 75.00 & 1.00 & 77.50 & 0.90 & 92.50 & 1.00 & 90.00 & 1.00 \\
16 & \texttt{Mistral-nemo:12b} & 50.00 & 1.30 & 57.50 & 1.10 & 67.50 & 1.10 & 90.00 & 1.40 & 97.50 & 1.50 \\
17 & \texttt{Mistral:7b} & 55.00 & 1.30 & 57.50 & 0.80 & 55.00 & 0.80 & 70.00 & 0.90 & 77.50 & 1.10 \\
18 & \texttt{Openchat:7b} & 15.38 & 1.30 & 65.00 & 0.80 & 67.50 & 0.80 & 82.50 & 1.00 & 85.00 & 1.00 \\
19 & \texttt{Orca-mini:3b} & 0.00 & 1.70 & 27.50 & 1.10 & 15.38 & 0.90 & 14.71 & 0.90 & 0.00 & 1.00 \\
20 & \texttt{Phi:7b} & 2.63 & 0.90 & 10.26 & 0.60 & 27.50 & 0.70 & 43.59 & 0.80 & 37.50 & 0.70 \\
21 & \texttt{Phi3:8b} & 0.00 & 1.40 & 29.73 & 0.90 & 43.24 & 0.80 & 47.37 & 0.80 & 52.63 & 0.90 \\
22 & \texttt{Qwen:4b} & 0.00 & 1.40 & 17.50 & 0.90 & 23.68 & 1.00 & 12.82 & 1.00 & 7.89 & 1.30 \\
23 & \texttt{Qwen2:7b} & 70.00 & 1.00 & 77.50 & 0.80 & 75.00 & 0.80 & 92.50 & 0.90 & 100.00 & 1.00 \\
24 & \texttt{Qwen2.5:7b} & 22.50 & 0.90 & 70.00 & 0.90 & 77.50 & 0.90 & 90.00 & 1.00 & 95.00 & 1.10 \\
25 & \texttt{QwQ-abliterated:32b} & 62.50 & 3.10 & 75.00 & 3.00 & 82.50 & 2.90 & 82.50 & 3.00 & 90.00 & 3.40 \\
26 & \texttt{QwQ-fusion:32b} & 57.50 & 3.00 & 75.00 & 3.00 & 77.50 & 2.90 & 87.50 & 3.10 & 90.00 & 3.40 \\
27 & \texttt{QwQ:32b} & 60.00 & 3.10 & 75.00 & 3.10 & 82.50 & 3.10 & 85.00 & 3.10 & 87.50 & 3.40 \\
28 & \texttt{Starcoder:3b} & 0.00 & 1.50 & 0.00 & 1.50 & 0.00 & 1.70 & 0.00 & 1.40 & 0.00 & 1.10 \\
29 & \texttt{Starcoder2:3b} & 0.00 & 0.40 & 0.00 & 0.40 & 0.00 & 0.40 & 0.00 & 0.50 & 0.00 & 0.40 \\
30 & \texttt{TinyLlama:1.1b} & 0.00 & 0.70 & 17.95 & 0.60 & 15.00 & 0.40 & 20.51 & 0.50 & 17.50 & 0.50 \\
31 & \texttt{Wizardlm2:7b} & 42.11 & 1.50 & 70.00 & 0.90 & 70.00 & 0.90 & 75.00 & 1.00 & 85.00 & 1.10 \\
32 & \texttt{Yi:6b} & 17.50 & 1.20 & 60.00 & 1.00 & 60.00 & 1.10 & 68.42 & 1.00 & 66.67 & 1.10 \\
33 & \texttt{Zephyr:7b} & 37.50 & 1.40 & 65.00 & 1.00 & 65.00 & 1.00 & 77.50 & 1.00 & 87.18 & 1.10 \\
\bottomrule
\end{tabular}%
}
\end{table*}

The inclusion of context examples significantly boosts performance for mid-range models like Codellama:7b and Dolphin-Mistral:7b, where accuracy improves up to 30\% as context increases from 0 to 9 examples. This trend highlights the importance of providing relevant examples to guide LLMs in understanding intent-specific nuances, especially for models optimized for in-context learning. However, some models, such as QwQ-fusion:32b, exhibit diminishing accuracy gains beyond a certain number of examples, suggesting a saturation point in leveraging additional context. Notably, specialized models like Deepseek-coder-v2:16b perform competitively, achieving 95\% accuracy with relatively low runtimes, showcasing the impact of domain-specific optimization. On the other hand, general-purpose models like Llama3.2:3b show steady but limited improvements, indicating a need for fine-tuning to handle domain-specific tasks effectively. In table \ref{Translation_NFV_best} we present the best LLMs for NFV configuration translation with respect to different context example.

\renewcommand{\arraystretch}{0.6}
\begin{table}[t!]
\caption{Best LLMs for intent translation for NFV configuration dataset}
\label{Translation_NFV_best}
\centering
\small
\resizebox{\columnwidth}{!}{%
\begin{tabular}{clcc}
\toprule
\textbf{Context} & \textbf{LLM} & \textbf{Accuracy (\%)} & \textbf{Avg. Time (s)} \\
\midrule
0 & \texttt{Codegemma:7b} & 75.0 & 1.3 \\
1 & \texttt{Command-r:35b} & 90.0 & 3.2 \\
3 & \texttt{Command-r:35b} & 87.5 & 3.5 \\
6 & \texttt{Command-r:35b} & 97.5 & 3.7 \\
9 & \texttt{Command-r:35b} & 100.0 & 4.0 \\
  & \texttt{Qwen2:7b}      & 100.0 & 1.0 \\
\bottomrule
\end{tabular}%
}
\end{table}

\noindent
\textbf{Intent2Flow-ODL Dataset: } Now we present the outcome of benchmarking using the proposed Intent2Flow-ODL dataset. This dataset evaluates ODL flow rule translation. For testing the LLMs translation capability, we used different natural language intents that use different network configuration tasks such as port based forwarding, firewall, flowspace slicing, IP based forwarding etc. Same as before, half of the dataset samples were used to give relevant examples to the LLM and the other half was used as test cases. We used three different values of context examples: 0, 1 and 3. Notably, on this dataset, we were able to get readable output from 30 LLMs out of targeted 33. Some LLMs failed to produce meaningful JSONs such as tinyLlama:1.1b, Orca-mini:3b and Deepseek-coder-v2:16b. The problem with Deepseek-coder-v2:16b was same as with Formal specification dataset, 'K-shift' not supported. As for tinyLlama:1.1b and Orca-mini:3b, the length of the designed prompt might cause the issue due to their context length limitation. We present the experimental results in Table \ref{tab:Translation_ODL}.

\renewcommand{\arraystretch}{0.5}
\begin{table}[t!]
\caption{Benchmarking of LLMs for intent translation for Intent2Flow-ODL dataset (\textbf{Ctx n} denotes n context examples)}
\label{tab:Translation_ODL}
\centering
\normalsize
\resizebox{\columnwidth}{!}{%
\begin{tabular}{ll|cc|cc|cc}
\toprule
\textbf{Sl.} & \textbf{LLM} & \multicolumn{2}{c|}{\textbf{Ctx 0}} & \multicolumn{2}{c|}{\textbf{Ctx 1}} & \multicolumn{2}{c}{\textbf{Ctx 3}} \\
\cmidrule(lr){3-4} \cmidrule(lr){5-6} \cmidrule(l){7-8}
& & Acc. & Time & Acc. & Time & Acc. & Time \\
\midrule
1 & \texttt{Codegemma:7b} & 0.00 & 2.42 & 88.89 & 3.11 & 77.78 & 3.20 \\
2 & \texttt{Codellama:34b} & 38.89 & 7.88 & 77.78 & 7.99 & 77.78 & 8.64 \\
3 & \texttt{Codellama:7b} & 27.78 & 2.39 & 77.78 & 2.23 & 83.33 & 2.40 \\
4 & \texttt{Codestral:22b} & 66.67 & 7.43 & 88.89 & 6.14 & 94.44 & 6.72 \\
5 & \texttt{Command-r:35b} & 27.78 & 6.99 & 88.89 & 7.20 & 100.00 & 7.50 \\
6 & \texttt{Deepseek-coder:1.3b} & 0.00 & 0.80 & 0.00 & 1.30 & 0.00 & 1.03 \\
7 & \texttt{Dolphin-Mistral:7b} & 0.00 & 2.67 & 55.56 & 2.53 & 77.78 & 2.24 \\
8 & \texttt{Gemma2:27b} & 50.00 & 6.04 & 88.89 & 6.01 & 100.00 & 5.66 \\
9 & \texttt{Llama2:7b} & 0.00 & 2.80 & 38.89 & 2.76 & 22.22 & 3.12 \\
10 & \texttt{Llama3:8b} & 33.33 & 2.54 & 94.44 & 2.18 & 83.33 & 1.98 \\
11 & \texttt{Llama3.1:8b} & 22.22 & 2.81 & 55.56 & 2.95 & 88.89 & 2.17 \\
12 & \texttt{Llama3.2:3b} & 0.00 & 1.51 & 50.00 & 1.17 & 72.22 & 1.30 \\
13 & \texttt{llava-Llama3:8b} & 0.00 & 1.85 & 11.11 & 2.06 & 33.33 & 1.99 \\
14 & \texttt{Marco-o1:7b} & 33.33 & 2.29 & 72.22 & 2.55 & 61.11 & 2.77 \\
15 & \texttt{Mistral-nemo:12b} & 50.00 & 3.11 & 55.56 & 2.08 & 77.78 & 2.42 \\
16 & \texttt{Mistral:7b} & 0.00 & 2.84 & 66.67 & 2.01 & 72.22 & 2.05 \\
17 & \texttt{Openchat:7b} & 0.00 & 2.12 & 72.22 & 1.82 & 83.33 & 1.96 \\
18 & \texttt{Phi:7b} & 0.00 & 0.98 & 0.00 & 1.59 & 0.00 & 0.51 \\
19 & \texttt{Phi3:8b} & 0.00 & 1.54 & 5.56 & 2.10 & 0.00 & 1.98 \\
20 & \texttt{Qwen:4b} & 0.00 & 0.81 & 0.00 & 1.30 & 0.00 & 3.15 \\
21 & \texttt{Qwen2:7b} & 38.89 & 2.25 & 61.11 & 1.76 & 83.33 & 1.90 \\
22 & \texttt{Qwen2.5:7b} & 44.44 & 2.23 & 83.33 & 2.11 & 72.22 & 2.73 \\
23 & \texttt{QwQ-abliterated:32b} & 83.33 & 8.29 & 94.44 & 6.51 & 100.00 & 6.31 \\
24 & \texttt{QwQ-fusion:32b} & 72.22 & 8.33 & 94.44 & 6.97 & 100.00 & 6.60 \\
25 & \texttt{QwQ:32b} & 83.33 & 8.26 & 94.44 & 6.98 & 100.00 & 6.25 \\
26 & \texttt{Starcoder:3b} & 0.00 & 0.34 & 0.00 & 0.44 & 0.00 & 0.44 \\
27 & \texttt{Starcoder2:3b} & 0.00 & 0.35 & 0.00 & 0.43 & 0.00 & 0.47 \\
28 & \texttt{Wizardlm2:7b} & 0.00 & 2.67 & 61.11 & 1.87 & 77.78 & 2.09 \\
29 & \texttt{Yi:6b} & 11.11 & 2.49 & 61.11 & 2.48 & 61.11 & 2.66 \\
30 & \texttt{Zephyr:7b} & 0.00 & 2.37 & 50.00 & 2.75 & 55.56 & 2.93 \\
\bottomrule
\end{tabular}%
}
\end{table}

The result is consistent with previous results from the Formal specification and NFV configuration datasets, larger models like QwQ-abliterated (32b) and QwQ-fusion (32b) achieve the highest accuracy (100\% with three context examples) due to their ability to generalize complex patterns in the dataset. These models, however, exhibit significantly higher runtimes (over 6 seconds on average), reflecting the computational demands of their large parameter size. Models such as Command-r (35b) and Gemma2 (27b) also perform well, demonstrating accuracy improvements of up to 70\% when transitioning from zero to three context examples, highlighting the importance of providing relevant examples to enhance in-context learning.

\renewcommand{\arraystretch}{0.6}
\begin{table}[t!]
\caption{Best LLMs for intent translation for Intent2Flow-ODL dataset}
\label{Translation_ODL_best}
\centering
\small
\resizebox{\columnwidth}{!}{%
\begin{tabular}{clcc}
\toprule
\textbf{Context} & \textbf{LLM} & \textbf{Accuracy (\%)} & \textbf{Avg. Time (s)} \\
\midrule
\multirow{2}{*}{0} 
    & \texttt{QwQ-abliterated:32b} & \multirow{2}{*}{83.33} & 8.29 \\
    & \texttt{QwQ:32b}             &                         & 8.26 \\
\midrule
\multirow{4}{*}{1} 
    & \texttt{Llama3:8b}           & \multirow{4}{*}{94.44} & 2.18 \\
    & \texttt{QwQ-abliterated:32b} &                        & 6.51 \\
    & \texttt{QwQ-fusion:32b}      &                        & 6.97 \\
    & \texttt{QwQ:32b}             &                        & 6.98 \\
\midrule
\multirow{5}{*}{3} 
    & \texttt{Command-r:35b}       & \multirow{5}{*}{100.00} & 7.50 \\
    & \texttt{Gemma2:27b}          &                         & 5.66 \\
    & \texttt{QwQ-abliterated:32b} &                         & 6.31 \\
    & \texttt{QwQ-fusion:32b}      &                         & 6.60 \\
    & \texttt{QwQ:32b}             &                         & 6.25 \\
\bottomrule
\end{tabular}%
}
\end{table}

\noindent
Mid-range models like Codegemma:7b and Codellama:7b show moderate improvements with added context but generally struggle to match the precision of larger models. Interestingly, Codellama:34b, despite its large size, exhibits less pronounced gains compared to QwQ-family models, suggesting that model specialization, in addition to parameter size, plays a critical role in handling domain-specific tasks like ODL flow rules. Consistent with observations from previous datasets, larger models are well-suited for complex intent translation tasks, but their high runtime makes them resource-intensive. The role of context examples is pivotal across all datasets, with significant accuracy improvements observed as examples increase, particularly for mid-sized models. In table \ref{Translation_ODL_best} we present the best LLMs for ODL flow rule translation with respect to different context example.

\begin{tcolorbox}[colback=gray!10!white, colframe=gray!80!black, 
  boxrule=0.5pt, arc=4pt, auto outer arc,
  left=6pt, right=6pt, top=6pt, bottom=6pt]
\textbf{Highlights:} 
On the Intent2Flow-ODL dataset, models like QwQ-abliterated and Command-r achieve very high accuracy with just three examples. Notably, the structured, rule-based format—enhanced by prompt decomposition (especially for QoS intents)—allows even mid-sized models like Codellama:7b and Llama3:8b to make substantial gains. This demonstrates that modular, intent-specific prompts can enable accurate SDN rule synthesis.
\end{tcolorbox}

\vspace{-0.15cm}

\noindent
\textbf{Intent2Flow-ONOS Dataset: } Now we present the outcome of benchmarking using the proposed Intent2Flow-ONOS dataset. We developed this dataset to evaluate LLMs on ONOS flow rule translation task. For testing the LLMs translation capability, we used diverse natural language intents that use different network configuration tasks such as port/IP based forwarding, blocking, flowspace slicing for QoS etc. Half of the dataset samples were used to give relevant examples to the LLM and the other half was used as test cases. We used five different values of context examples: 0, 1, 3, 6 and 9. Similar to Intent2Flow-ODL dataset, we were able to get readable output from 30 LLMs out of targeted 33. TinyLlama:1.1b, Orca-mini:3b and Deepseek-coder-v2:16b failed to produce meaningful JSON structures. We present the experimental results in Table \ref{tab:Translation_ONOS}.

\renewcommand{\arraystretch}{0.6}
\begin{table*}[t!]
\caption{Benchmarking of LLMs for intent translation for Intent2Flow-ONOS dataset (\textbf{Ctx n} denotes n context examples)}
\label{tab:Translation_ONOS}
\centering
\tiny
\resizebox{\textwidth}{!}{%
\begin{tabular}{ll|cc|cc|cc|cc|cc}
\toprule
\textbf{Sl.} & \textbf{LLM} & \multicolumn{2}{c|}{\textbf{Ctx 0}} & \multicolumn{2}{c|}{\textbf{Ctx 1}} & \multicolumn{2}{c|}{\textbf{Ctx 3}} & \multicolumn{2}{c|}{\textbf{Ctx 6}} & \multicolumn{2}{c}{\textbf{Ctx 9}} \\
\cmidrule(lr){3-4} \cmidrule(lr){5-6} \cmidrule(lr){7-8} \cmidrule(lr){9-10} \cmidrule(l){11-12}
& & Acc. & Time & Acc. & Time & Acc. & Time & Acc. & Time & Acc. & Time \\
\midrule
1 & \texttt{Codegemma:7b} & 36 & 2.25 & 84 & 2.71 & 92 & 2.74 & 92 & 2.56 & 88 & 2.66 \\
2 & \texttt{Codellama:34b} & 48 & 6.92 & 72 & 7.03 & 72 & 8.09 & 72 & 9.26 & 64 & 10.26 \\
3 & \texttt{Codellama:7b} & 40 & 1.95 & 68 & 1.59 & 80 & 1.68 & 80 & 1.99 & 88 & 2.22 \\
4 & \texttt{Codestral:22b} & 96 & 5.93 & 96 & 3.59 & 100 & 4.15 & 100 & 4.57 & 100 & 5.14 \\
5 & \texttt{Command-r:35b} & 92 & 5.77 & 92 & 6.39 & 96 & 6.68 & 100 & 7.13 & 96 & 7.68 \\
6 & \texttt{Deepseek-coder:1.3b} & 0 & 1.25 & 0 & 1.46 & 0 & 1.84 & 0 & 1.66 & 0 & 1.89 \\
7 & \texttt{Dolphin-Mistral:7b} & 0 & 2.36 & 40 & 2.48 & 60 & 2.98 & 68 & 3.17 & 80 & 2.82 \\
8 & \texttt{Gemma2:27b} & 88 & 6.06 & 92 & 3.95 & 88 & 4.08 & 92 & 4.49 & 92 & 5.00 \\
9 & \texttt{Llama2:7b} & 4 & 2.51 & 32 & 2.45 & 20 & 2.88 & 8 & 3.20 & 4 & 3.40 \\
10 & \texttt{Llama3.1:8b} & 0 & 2.06 & 64 & 2.10 & 72 & 2.11 & 88 & 1.73 & 84 & 2.01 \\
11 & \texttt{Llama3.2:3b} & 24 & 1.24 & 40 & 0.94 & 52 & 1.14 & 60 & 0.95 & 64 & 1.05 \\
12 & \texttt{Llama3:8b} & 56 & 2.06 & 60 & 1.77 & 76 & 1.33 & 80 & 1.51 & 80 & 1.54 \\
13 & \texttt{llava-Llama3:8b} & 16 & 2.02 & 48 & 1.49 & 56 & 1.41 & 64 & 2.01 & 60 & 1.75 \\
14 & \texttt{Marco-o1:7b} & 68 & 1.80 & 88 & 2.09 & 92 & 2.20 & 84 & 1.84 & 80 & 1.98 \\
15 & \texttt{Mistral:7b} & 44 & 2.34 & 20 & 1.57 & 44 & 1.60 & 72 & 1.82 & 68 & 1.93 \\
16 & \texttt{Mistral-nemo:12b} & 80 & 2.65 & 88 & 2.60 & 92 & 2.59 & 96 & 2.74 & 92 & 3.32 \\
17 & \texttt{Openchat:7b} & 24 & 2.02 & 72 & 2.76 & 72 & 3.40 & 80 & 2.11 & 68 & 1.91 \\
18 & \texttt{Phi:7b} & 0 & 0.89 & 20 & 1.19 & 16 & 1.10 & 12 & 2.69 & 0 & 2.06 \\
19 & \texttt{Phi3:8b} & 0 & 1.69 & 4 & 1.61 & 4 & 1.86 & 16 & 1.79 & 20 & 1.93 \\
20 & \texttt{Qwen:4b} & 0 & 1.72 & 8 & 1.97 & 0 & 1.39 & 0 & 2.90 & 0 & 4.51 \\
21 & \texttt{Qwen2.5} & 68 & 1.86 & 88 & 1.80 & 84 & 1.56 & 84 & 1.59 & 88 & 1.68 \\
22 & \texttt{Qwen2:7b} & 40 & 1.66 & 64 & 1.60 & 80 & 1.33 & 88 & 1.46 & 88 & 1.60 \\
23 & \texttt{QwQ:32b} & 88 & 6.80 & 80 & 7.04 & 92 & 5.92 & 96 & 5.74 & 96 & 6.06 \\
24 & \texttt{QwQ-abliterated:32b} & 80 & 6.82 & 88 & 5.79 & 96 & 4.84 & 100 & 5.21 & 96 & 5.77 \\
25 & \texttt{QwQ-fusion:32b} & 80 & 6.61 & 84 & 6.51 & 80 & 5.32 & 96 & 5.61 & 96 & 6.12 \\
26 & \texttt{Starcoder:3b} & 0 & 0.64 & 0 & 1.19 & 0 & 0.90 & 0 & 1.24 & 0 & 0.67 \\
27 & \texttt{Starcoder2:3b} & 0 & 0.45 & 0 & 0.45 & 0 & 0.65 & 0 & 0.70 & 0 & 0.63 \\
28 & \texttt{Wizardlm2:7b} & 36 & 2.48 & 44 & 2.49 & 52 & 2.23 & 52 & 2.77 & 60 & 3.22 \\
29 & \texttt{Yi:6b} & 36 & 2.20 & 40 & 2.00 & 52 & 1.87 & 48 & 2.08 & 64 & 2.46 \\
30 & \texttt{Zephyr:7b} & 20 & 2.13 & 60 & 2.30 & 44 & 2.52 & 56 & 2.83 & 60 & 3.10 \\
\bottomrule
\end{tabular}%
}
\end{table*}

Table \ref{tab:Translation_ONOS} reveals some clear trends. Several models, notably Codestral:22b, Command-r:35b, and various QwQ models (32b), achieved high accuracy, often exceeding 90\% and even reaching 100\% on some contexts. This suggests that for this specific task, certain architectures and training regimens are more effective. Interestingly, parameter size doesn't seem to be the sole determinant of success. While some larger models like Command-r:35b performed reasonably well, they were outperformed by smaller models like Codestral:22b. This indicates that models specializing in code generation and structured output translation, like those found in the ``code'' prefixed models (Codegemma, Codellama, Codestral), play a crucial role. The trade-off between accuracy and inference time was also evident, with larger models taking significantly longer—Codellama-34b, for instance, required over 10 seconds per inference, whereas Qwen2.5-7B achieved competitive performance in under 2 seconds. Conversely, models like Deepseek-coder, Phi, Qwen, and Starcoder variants struggled, often achieving 0\% accuracy, suggesting they might not be suitable for this type of translation task, perhaps being geared towards different NLP applications. The Llama family shows varied results, with some larger variants performing better than smaller ones, but still not reaching the top performers' level. This highlights the importance of not just scale, but also the data and training methodology. Furthermore, context sensitivity is also important as most models improved with more added examples. For instance, Dolphin-Mistral:7b jumped from 0\% to 80\%. Using 6–9 context examples is beneficial, as most models reach peak performance within this range.

\begin{tcolorbox}[colback=gray!10!white, colframe=gray!80!black, 
  boxrule=0.5pt, arc=4pt, auto outer arc,
  left=6pt, right=6pt, top=6pt, bottom=6pt]
\textbf{Highlights:}
In the ONOS flow rule translation task, structured prompts matching the ONOS JSON schema separated queue/VLAN intents from forwarding/blocking rules. Combined with rigid SDN-specific intents, this enabled Codestral:22b to achieve very high accuracy with low latency, outperforming larger LLMs. Specialized prompts, task regularity, and 6–9 in-context examples boosted performance, especially for mid-sized models like Qwen2.5:7b.
\end{tcolorbox}

Given these findings, for ONOS intent translation, models like Codestral:22b, Command-r:35b, and the QwQ family appear to be the most promising. For other SDN controllers, the lessons learned should be similar: prioritize models specialized in code generation or translation and don't solely rely on parameter count.  Benchmarking with representative intent examples is crucial for selecting the right LLM. Further investigation into the training data and architecture of high-performing models would be beneficial for developing even more effective solutions for intent translation in SDN controllers. Moreover, fine-tuning mid-sized model (e.g., Mistral-Nemo-12b or Qwen2.5-7b) could be a more practical choice, offering a trade off between speed and accuracy.
In table \ref{Translation_ONOS_best} we present the best LLMs for ONOS flow rule translation with respect to different context example.

\renewcommand{\arraystretch}{0.6}
\begin{table}[t!]
\caption{Best LLMs for intent translation for Intent2Flow-ONOS dataset}
\label{Translation_ONOS_best}
\centering
\small
\resizebox{\columnwidth}{!}{%
\begin{tabular}{clcc}
\toprule
\textbf{Context} & \textbf{LLM} & \textbf{Accuracy (\%)} & \textbf{Avg. Time (s)} \\
\midrule
0 & \texttt{Codestral:22b} & 96 & 5.93 \\
1 & \texttt{Codestral:22b} & 96 & 3.59 \\
3 & \texttt{Codestral:22b} & 100 & 4.15 \\
\midrule
\multirow{3}{*}{6} 
  & \texttt{Codestral:22b}           & \multirow{3}{*}{100} & 4.57 \\
  & \texttt{Command-r:35b}           &                       & 7.13 \\
  & \texttt{QwQ-abliterated:32b}     &                       & 5.21 \\
\midrule
9 & \texttt{Codestral:22b} & 100 & 5.14 \\
\bottomrule
\end{tabular}%
}
\end{table}

\textbf{Evaluation of 70 Billion Parameter LLMs:} Our main benchmarking focused on small to mid-sized LLMs, leaving out models at the extreme end of the parameter scale. To understand how very large models perform on the same task, we evaluated three representative 70-billion-parameter LLMs on the Intent2Flow-ONOS dataset. Their results, shown in Table~\ref{tab:translation_ONOS_70b}, provide insight into the capabilities and limitations of scaling up model size for SDN intent translation. 

\renewcommand{\arraystretch}{0.6}
\begin{table*}[t!]
\caption{Benchmarking of 70 billion parameter LLMs for intent translation for Intent2Flow-ONOS dataset (\textbf{Ctx n} denotes n context examples)}
\label{tab:translation_ONOS_70b}
\centering
{%

\fontsize{3.5}{5.5}\selectfont  

\resizebox{\textwidth}{!}{%
\begin{tabular}{ll|cc|cc|cc|cc|cc}
\toprule
\textbf{Sl.} & \textbf{LLM} & \multicolumn{2}{c|}{\textbf{Ctx 0}} & \multicolumn{2}{c|}{\textbf{Ctx 1}} & \multicolumn{2}{c|}{\textbf{Ctx 3}} & \multicolumn{2}{c|}{\textbf{Ctx 6}} & \multicolumn{2}{c}{\textbf{Ctx 9}} \\
\cmidrule(lr){3-4} \cmidrule(lr){5-6} \cmidrule(lr){7-8} \cmidrule(lr){9-10} \cmidrule(l){11-12}
& & Acc. & Time & Acc. & Time & Acc. & Time & Acc. & Time & Acc. & Time \\
\midrule
1 & \texttt{Codellama:70b} & 60 & 14.73 & 48 & 10.45 & 60 & 11.89 & 72 & 12.37 & 68 & 15.50 \\
2 & \texttt{Llama2:70b}    & 8  & 16.26 & 64 & 14.08 & 72 & 14.89 & 76 & 15.77 & 76 & 17.18 \\
3 & \texttt{Llama3.3:70b}  & 88 & 10.43 &    &       &    &       &    &       &    &       \\
\bottomrule
\end{tabular}%
}}
\end{table*}

The evaluation reveals that larger model size does not inherently guarantee superior performance in translating high-level network intents into ONOS-compatible flow rules. Notably, Llama3.3:70b achieved an 88\% accuracy with zero-shot prompts, surpassing both Llama2:70b and Codellama:70b. We could not collect results for larger context examples for Llama 3.3 due to hardware memory constraints. Codellama:70b and Llama2:70b showed only modest accuracy gains despite increasing context, and their overall performance lagged behind smaller, code-specialized models like Codestral:22b. This limited improvement may stem from a mismatch between the Intent2Flow-ONOS dataset (which is structurally rigid and highly schema-driven) and the broader training objectives of general-purpose 70b models. Additionally, the multi-prompt strategy used in our benchmarking was tuned toward compact, code-oriented models; larger models not specifically fine-tuned for structured translation tasks may struggle to align with such narrow formats. These discrepancies suggest that the effectiveness of large models is heavily influenced by prompt design and the nature of the dataset. 

\begin{tcolorbox}[colback=gray!10!white, colframe=gray!80!black, 
  boxrule=0.5pt, arc=4pt, auto outer arc,
  left=6pt, right=6pt, top=6pt, bottom=6pt]
\textbf{Highlights:} Despite their size, 70B models like Codellama and Llama2 delivered limited gains on the Intent2Flow-ONOS task, likely due to a mismatch between their general-purpose training and the dataset’s rigid, schema-driven structure—emphasizing that effective intent translation depends more on prompt-task alignment than on model scale alone.
\end{tcolorbox}

Moreover, this underscores the importance of tailoring prompts to the specific capabilities of the model and the characteristics of the dataset. In contrast, smaller models have demonstrated more consistent and reliable performance which highlights that model size should be considered alongside other factors such as prompt engineering and dataset alignment when evaluating LLMs for IBN tasks.

\noindent
\textbf{Summary of Benchmarking for Intent Translation:} The benchmarking of LLMs across four datasets—Formal Specification, NFV Configuration, Intent2Flow-ODL, and Intent2Flow-ONOS—reveals key trends in performance, efficiency, and generalization. Larger models such as the QwQ family (32b) and Command-r (35b) consistently achieved the highest accuracy (often $\geq 99\%$), particularly for structurally rigid tasks, though at the cost of higher runtimes. Mid-sized models like Codellama:7b and Dolphin-Mistral:7b showed substantial accuracy gains (up to 30\%) with more context examples, highlighting the strong impact of in-context learning when paired with well-designed prompts. Smaller models (e.g., TinyLlama:1.1b, Orca-mini:3b) struggled to generalize, often failing to produce syntactically valid outputs due to parameter and context limitations. Interestingly, the evaluation of 70B parameter LLMs (e.g., Llama2, Llama3.3, and Codellama:70b) showed that size alone does not ensure superiority: Codellama and Llama2 offered only marginal improvements over mid-sized models, and at significantly higher computational cost. Llama3.3:70b performed better in zero-shot settings, but further results were limited by hardware constraints. These findings suggest that beyond scale, alignment between the model’s training, prompt structure, and the schema-specific nature of intent translation tasks is critical for reliable performance. Moreover, diminishing gains for some large models (e.g., QwQ-fusion) with more context imply a saturation effect, reinforcing the importance of task-specialized tuning over raw parameter count.

Notably, benchmarking results show that LLMs generally perform better on ONOS than on ODL in zero-shot (Ctx 0) scenarios, with most models achieving higher accuracy without context examples. However, as more context is provided (Ctx 1 and Ctx 3), this advantage shifts, and a greater number of models achieve higher accuracy on ODL. This is likely because ONOS’s schema is simpler and more closely aligned with common pretraining data, enabling better zero-shot performance, while in-context examples help LLMs adapt to ODL’s more complex structure and narrow the performance gap.

\vspace{-0.3cm}
\subsection{LLM BENCHMARKING RESULTS FOR CONFLICT DETECTION}
To find how LLMs perform on conflict detection task related to network configuration, we benchmarked them on the proposed FlowConflict-ODL and FlowConflict-ONOS datasets.

\noindent
\textbf{Conflict Detection using FlowConflict-ODL Datasets: } 
We now present the results of benchmarking the LLMs for the conflict detection task using the FlowConflict-ODL dataset. In Sec. IV-\ref{FlowConflict_datasets}, we mentioned how the dataset is prepared. In total, each LLM evaluated 50 pairs of flow rules. Among these, 46 rule pairs were non-conflicting, while 4 were conflicting. Therefore, the ideal outcomes are 4 true positives (TP) and 46 true negatives (TN). Table~\ref{tab:conflict_ODL} presents the TP, TN, false positives (FP), and false negatives (FN) for 28 LLMs. We excluded 5 models (``Phi", ``Orca-mini", ``Qwen", ``tinyLlama", ``Deepseek-coder-V2") from the results due to their inability to generate meaningful responses.

\renewcommand{\arraystretch}{0.6}
\begin{table*}[t!]
\centering
\caption{Benchmarking of LLMs for conflict detection using FlowConflict-ODL dataset}
\label{tab:conflict_ODL}
\tiny
\resizebox{\textwidth}{!}{%
\begin{tabular}{llcccccccccc}
\toprule
\textbf{Sl.} & \textbf{Model} & \textbf{TP} & \textbf{FP} & \textbf{FN} & \textbf{TN} & \textbf{Accuracy} & \textbf{Precision} & \textbf{Recall} & \textbf{F1-Score} & \textbf{FPR} & \textbf{Avg. Time} \\
\midrule
1 & \texttt{Marco-o1:7b} & 1 & 12 & 3 & 34 & 0.70 & 0.08 & 0.25 & 0.12 & 0.26 & 4.19 \\
2 & \texttt{Mistral:7b} & 4 & 46 & 0 & 0 & 0.08 & 0.08 & 1.00 & 0.15 & 1.00 & 4.05 \\
3 & \texttt{Mistral-nemo:12b} & 4 & 39 & 0 & 7 & 0.22 & 0.09 & 1.00 & 0.17 & 0.85 & 5.61 \\
4 & \texttt{Deepseek-coder:1.3b} & 3 & 43 & 1 & 3 & 0.12 & 0.07 & 0.75 & 0.12 & 0.93 & 2.58 \\
5 & \texttt{Starcoder:3b} & 0 & 0 & 4 & 46 & 0.92 & 0.00 & 0.00 & 0.00 & 0.00 & 3.85 \\
6 & \texttt{Codegemma:7b} & 4 & 22 & 0 & 24 & 0.56 & 0.15 & 1.00 & 0.27 & 0.48 & 4.09 \\
7 & \texttt{Starcoder2:3b} & 0 & 0 & 4 & 46 & 0.92 & 0.00 & 0.00 & 0.00 & 0.00 & 3.18 \\
8 & \texttt{Openchat:7b} & 3 & 21 & 1 & 25 & 0.56 & 0.12 & 0.75 & 0.21 & 0.46 & 3.48 \\
9 & \texttt{Phi3:8b} & 2 & 25 & 2 & 21 & 0.46 & 0.07 & 0.50 & 0.13 & 0.54 & 2.63 \\
10 & \texttt{Dolphin-Mistral:7b} & 2 & 19 & 2 & 27 & 0.58 & 0.10 & 0.50 & 0.16 & 0.41 & 3.39 \\
11 & \texttt{Wizardlm2:7b} & 4 & 46 & 0 & 0 & 0.08 & 0.08 & 1.00 & 0.15 & 1.00 & 4.09 \\
12 & \texttt{Yi:6b} & 4 & 46 & 0 & 0 & 0.08 & 0.08 & 1.00 & 0.15 & 1.00 & 3.61 \\
13 & \texttt{Zephyr:7b} & 2 & 26 & 2 & 20 & 0.44 & 0.07 & 0.50 & 0.12 & 0.57 & 3.70 \\
14 & \texttt{Command-r:35b} & 2 & 11 & 2 & 35 & 0.74 & 0.15 & 0.50 & 0.24 & 0.24 & 9.08 \\
15 & \texttt{llava-Llama3:8b} & 1 & 16 & 3 & 30 & 0.62 & 0.06 & 0.25 & 0.10 & 0.35 & 4.71 \\
16 & \texttt{Codestral:22b} & 2 & 8 & 2 & 38 & 0.80 & 0.20 & 0.50 & 0.29 & 0.17 & 7.14 \\
17 & \texttt{Codellama:34b} & 3 & 29 & 1 & 17 & 0.40 & 0.09 & 0.75 & 0.17 & 0.63 & 3.87 \\
18 & \texttt{Codellama:7b} & 4 & 45 & 0 & 1 & 0.10 & 0.08 & 1.00 & 0.15 & 0.98 & 3.87 \\
19 & \texttt{Llama2:7b} & 2 & 30 & 2 & 16 & 0.36 & 0.06 & 0.50 & 0.11 & 0.65 & 3.69 \\
20 & \texttt{Llama3:8b} & 4 & 36 & 0 & 10 & 0.28 & 0.10 & 1.00 & 0.18 & 0.78 & 4.27 \\
21 & \texttt{Llama3.1:8b} & 2 & 22 & 2 & 24 & 0.52 & 0.08 & 0.50 & 0.14 & 0.48 & 4.25 \\
22 & \texttt{Llama3.2:3b} & 2 & 26 & 2 & 20 & 0.44 & 0.07 & 0.50 & 0.12 & 0.57 & 3.27 \\
23 & \texttt{Qwen2:7b} & 2 & 17 & 2 & 29 & 0.62 & 0.11 & 0.50 & 0.17 & 0.37 & 3.52 \\
24 & \texttt{Qwen2.5:7b} & 0 & 3 & 4 & 43 & 0.86 & 0.00 & 0.00 & 0.00 & 0.07 & 3.59 \\
25 & \texttt{Gemma2:27b} & 2 & 7 & 2 & 39 & 0.82 & 0.22 & 0.50 & 0.31 & 0.15 & 7.72 \\
26 & \texttt{QwQ-abliterated:32b} & 4 & 4 & 0 & 42 & 0.92 & 0.50 & 1.00 & 0.67 & 0.09 & 9.23 \\
27 & \texttt{QwQ-fusion:32b} & 4 & 7 & 0 & 39 & 0.86 & 0.36 & 1.00 & 0.53 & 0.15 & 9.59 \\
28 & \texttt{QwQ:32b} & 3 & 4 & 1 & 42 & 0.90 & 0.43 & 0.75 & 0.55 & 0.09 & 9.42 \\
\bottomrule
\end{tabular}%
}
\end{table*}

The results reveal significant variations in accuracy, precision, and error handling across models. Models such as QwQ-abliterated:32b, QwQ-fusion:32b demonstrated exemplary performance, achieving near-perfect results (TP = 4, FP around 7, TN around 39) with minimal FP or FN. QwQ fell slightly behind by achieving TP = 3. These results are consistent with the earlier findings, where the QwQ-family models consistently excelled in both natural language and JSON-based conflict detection tasks, indicating their strong generalization and task-specific capabilities. Anomalies arise with models like Mistral:7b, Wizardlm2:7b, and Yi:6b, which produced perfect True Positives (TP = 4) but had excessively high false positives (FP = 46), suggesting that these models lacked the ability to correctly differentiate conflicting from non-conflicting cases. This pattern is inconsistent with earlier evaluations, where Mistral:7b and Wizardlm2:7b performed moderately well, indicating sensitivity to the input format (natural language vs. JSON). Conversely, models such as Starcoder:3b and Starcoder2:3b failed entirely (TP = 0, FN = 4), reflecting their inability to interpret the conflict detection task in the structured JSON format, consistent with their poor performance in earlier JSON-related evaluations. Larger models like Codestral:22b and Gemma2:27b displayed mixed results, with moderate TPs (2) but noticeable FPs (7–8). This suggests that while they can identify conflicts to some extent, they are prone to over-flagging non-conflicting cases. These results align with prior observations, where these models showed decent, but not exceptional, performance in JSON-based tasks.

The performance of Codellama:34b is notably underwhelming, especially given its large parameter size and computational demands. With a True Positive (TP) count of 3, False Positives (FP) of 29, and a True Negative (TN) of only 17, it performed worse than similar and mid-sized models like Codestral:22b, Codegemma:7b and Command-r:35b. This discrepancy indicates that Codellama:34b, despite its size and potential for handling complex tasks, struggles to balance sensitivity and precision in JSON-based conflict detection. A likely reason for this underperformance is the model's inability to effectively handle structured data, as seen in earlier results where it also lagged behind in JSON-based tasks. This could stem from inadequate pretraining or fine-tuning on tasks involving structured formats like JSON. Furthermore, its high FP count suggests that it frequently over-flagged non-conflicting cases as conflicts, reflecting poor understanding of the nuanced prompt instructions.

\begin{tcolorbox}[colback=gray!10!white, colframe=gray!80!black, 
  boxrule=0.5pt, arc=4pt, auto outer arc,
  left=6pt, right=6pt, top=6pt, bottom=6pt]
\textbf{Highlights:} 
In the FlowConflict-ODL conflict detection task, a strict field-by-field comparison prompt led to varied LLM performance. Only QwQ-abliterated and Gemma2:27b accurately detected direct conflicts with minimal false positives, while others over-predicted due to partial match misinterpretation or missing-field semantics errors. Reliable detection requires precise structural reasoning aligned with literal matching rules, achieved by few code-aligned or structurally sensitive models.
\end{tcolorbox}

\vspace{-0.3cm}
\noindent
\textbf{Conflict Detection using FlowConflict-ONOS Datasets.: } We now present the results of benchmarking the LLMs for the conflict detection task using the FlowConflict-ONOS dataset. In Sec. IV-\ref{FlowConflict_datasets}, we mentioned how the dataset is prepared. In total, each LLM evaluated 62 pairs of flow rules. Among these, 52 rules were non-conflicting, while 10 were conflicting. Therefore, the ideal outcomes are 10 true positives (TP) and 52 true negatives (TN). Table~\ref{tab:Conflict_ONOS} presents the TP, TN, FP, and FN for 30 LLMs. We excluded 3 models (``Orca-mini", ``tinyLlama", ``Deepseek-coder-V2") from the results due to their inability to generate meaningful responses.

The evaluation reveals significant performance variations among LLMs in detecting conflicting ONOS flow rules. QwQ-32b stands out as the only model achieving perfect accuracy (100\%), with zero false positives or false negatives, making it the gold standard for this task. Close behind, QwQ-fusion-32b (98\%) and QwQ-abliterated-32b (94\%) also demonstrate exceptional performance, suggesting that the QwQ family is particularly well-suited for structured data conflict identification.

\renewcommand{\arraystretch}{0.6}
\begin{table*}[t!]
\caption{Benchmarking of LLMs for conflict detection using FlowConflict-ONOS dataset}
\label{tab:Conflict_ONOS}
\centering
\tiny
\resizebox{\textwidth}{!}{%
\begin{tabular}{llcccccccccc}
\toprule
\textbf{Sl.} & \textbf{Model} & \textbf{TP} & \textbf{FP} & \textbf{TN} & \textbf{FN} & \textbf{Accuracy} & \textbf{Precision} & \textbf{Recall} & \textbf{F1-Score} & \textbf{FPR} & \textbf{Avg. Time} \\
\midrule
1 & \texttt{Codegemma:7b} & 5 & 43 & 9 & 5 & 0.23 & 0.10 & 0.50 & 0.17 & 0.83 & 21.67 \\
2 & \texttt{Codellama:34b} & 9 & 47 & 5 & 1 & 0.23 & 0.16 & 0.90 & 0.27 & 0.90 & 52.27 \\
3 & \texttt{Codellama:7b} & 10 & 52 & 0 & 0 & 0.16 & 0.16 & 1.00 & 0.28 & 1.00 & 18.83 \\
4 & \texttt{Codestral:22b} & 9 & 30 & 22 & 1 & 0.50 & 0.23 & 0.90 & 0.37 & 0.58 & 37.31 \\
5 & \texttt{Command-r:35b} & 8 & 15 & 37 & 2 & 0.73 & 0.35 & 0.80 & 0.49 & 0.29 & 46.75 \\
6 & \texttt{Deepseek-coder:1.3b} & 9 & 49 & 3 & 1 & 0.19 & 0.16 & 0.90 & 0.27 & 0.94 & 10.57 \\
7 & \texttt{Dolphin-Mistral:7b} & 9 & 45 & 7 & 1 & 0.26 & 0.17 & 0.90 & 0.29 & 0.87 & 16.55 \\
8 & \texttt{Gemma2:27b} & 8 & 3 & 49 & 2 & 0.92 & 0.73 & 0.80 & 0.76 & 0.06 & 38.19 \\
9 & \texttt{Llama2:7b} & 10 & 52 & 0 & 0 & 0.16 & 0.16 & 1.00 & 0.28 & 1.00 & 17.72 \\
10 & \texttt{Llama3.1:8b} & 3 & 22 & 30 & 7 & 0.53 & 0.12 & 0.30 & 0.17 & 0.42 & 20.38 \\
11 & \texttt{Llama3.2:3b} & 5 & 26 & 26 & 5 & 0.50 & 0.16 & 0.50 & 0.24 & 0.50 & 16.50 \\
12 & \texttt{Llama3:8b} & 6 & 36 & 16 & 4 & 0.35 & 0.14 & 0.60 & 0.23 & 0.69 & 20.95 \\
13 & \texttt{llava-Llama3:8b} & 4 & 30 & 22 & 6 & 0.42 & 0.12 & 0.40 & 0.18 & 0.58 & 23.96 \\
14 & \texttt{Marco-o1:7b} & 6 & 10 & 42 & 4 & 0.77 & 0.38 & 0.60 & 0.47 & 0.19 & 21.42 \\
15 & \texttt{Mistral:7b} & 8 & 51 & 1 & 2 & 0.15 & 0.14 & 0.80 & 0.24 & 0.98 & 17.12 \\
16 & \texttt{Mistral-nemo:12b} & 10 & 38 & 14 & 0 & 0.39 & 0.21 & 1.00 & 0.35 & 0.73 & 27.19 \\
17 & \texttt{Openchat:7b} & 9 & 49 & 3 & 1 & 0.19 & 0.16 & 0.90 & 0.27 & 0.94 & 17.76 \\
18 & \texttt{Phi:7b} & 7 & 43 & 9 & 3 & 0.26 & 0.14 & 0.70 & 0.23 & 0.83 & 11.87 \\
19 & \texttt{Phi3:8b} & 9 & 48 & 4 & 1 & 0.21 & 0.16 & 0.90 & 0.27 & 0.92 & 13.25 \\
20 & \texttt{Qwen:4b} & 1 & 1 & 51 & 9 & 0.84 & 0.50 & 0.10 & 0.17 & 0.02 & 8.01 \\
21 & \texttt{Qwen2.5:7b} & 3 & 1 & 51 & 7 & 0.87 & 0.75 & 0.30 & 0.43 & 0.02 & 19.10 \\
22 & \texttt{Qwen2:7b} & 5 & 8 & 44 & 5 & 0.79 & 0.38 & 0.50 & 0.43 & 0.15 & 19.42 \\
23 & \texttt{QwQ:32b} & 10 & 0 & 52 & 0 & 1.00 & 1.00 & 1.00 & 1.00 & 0.00 & 47.43 \\
24 & \texttt{QwQ-abliterated:32b} & 9 & 3 & 49 & 1 & 0.94 & 0.75 & 0.90 & 0.82 & 0.06 & 47.73 \\
25 & \texttt{QwQ-fusion:32b} & 10 & 1 & 51 & 0 & 0.98 & 0.91 & 1.00 & 0.95 & 0.02 & 47.53 \\
26 & \texttt{Starcoder:3b} & 0 & 0 & 52 & 10 & 0.84 & -- & 0.00 & -- & 0.00 & 15.30 \\
27 & \texttt{Starcoder2:3b} & 0 & 0 & 52 & 10 & 0.84 & -- & 0.00 & -- & 0.00 & 12.93 \\
28 & \texttt{Wizardlm2:7b} & 8 & 51 & 1 & 2 & 0.15 & 0.14 & 0.80 & 0.24 & 0.98 & 20.16 \\
29 & \texttt{Yi:6b} & 8 & 45 & 7 & 2 & 0.24 & 0.15 & 0.80 & 0.25 & 0.87 & 16.31 \\
30 & \texttt{Zephyr:7b} & 9 & 35 & 17 & 1 & 0.42 & 0.20 & 0.90 & 0.33 & 0.67 & 18.45 \\
\bottomrule
\end{tabular}%
}
\end{table*}

\begin{tcolorbox}[colback=gray!10!white, colframe=gray!80!black, 
  boxrule=0.5pt, arc=4pt, auto outer arc,
  left=6pt, right=6pt, top=6pt, bottom=6pt]
\textbf{Highlights:}
In the FlowConflict-ONOS conflict detection task, QwQ:32b uniquely achieved very high accuracy with no false positives. The strict, schema-based prompt required exact field-by-field flow rule comparison, revealing weaknesses in models using generalization or fuzzy matching. Precise conflict detection demands structural alignment between the LLM’s behavior and the dataset’s deterministic logic, which only highly structured-output models like QwQ:32b consistently achieved.
\end{tcolorbox}

Llama3.3-70b (94\%) and Gemma2-27b (92\%) also perform well, but at significantly higher computational costs. Many models, especially some of the smaller ones or those not specifically designed for code-related tasks, struggled significantly. For instance, several models, including Llama2:70b and Llama2:7b, had a high recall (correctly identifying most actual conflicts) but also an extremely high FPR, meaning they flagged many non-conflicts as conflicts, rendering them practically unfit. Similarly, many models struggle with false positives, with Codellama:7b, Codegemma-7b, Dolphin-Mistral-7b, and Deepseek-coder-1.3b exceeding 87\% false positive rate (FPR), making them unsuitable for practical use indicating limited contextual understanding of JSON flow rules. However, the Starcoder models, despite being code-focused, had very low recall. The same conclusion can be drawn here as was in intent translation- parameter size alone doesn't guarantee good performance, as some larger models performed poorly. The table reveals a clear speed-performance trade-off. While QwQ:32b delivers perfect performance, it's not the fastest model. Some smaller models, like those from the Phi family, are quicker but have much lower accuracy. Given these findings, for conflict detection in ONOS flow rules, the QwQ family is the clear recommendation balancing accuracy and efficiency. Prioritizing models with high precision (to minimize false alarms) and balanced F1-scores is critical, though deployment should consider inference time as well. As for SDN controllers in general, the key lesson is to prioritize models that have demonstrated strong performance on similar rule-based tasks or code analysis. Careful evaluation with representative flow rules and consideration of the speed-performance trade-off are essential. A model with high recall but also a very high false positive rate is not practical.

\vspace{-0.4cm}
\subsection{EVALUATION OF 70 BILLION PARAMETER LLMS FOR CONFLICT DETECTION}

\renewcommand{\arraystretch}{0.6}
\begin{table*}[t!]
\caption{Benchmarking of 70 billion parameter LLMs for conflict detection using FlowConflict-ONOS dataset}
\label{tab:conflict_ONOS_70b}
\centering
\tiny
\resizebox{\textwidth}{!}{%
\begin{tabular}{llcccccccccc}
\toprule
\textbf{Sl.} & \textbf{Model} & \textbf{TP} & \textbf{FP} & \textbf{TN} & \textbf{FN} & \textbf{Accuracy} & \textbf{Precision} & \textbf{Recall} & \textbf{F1-Score} & \textbf{FPR} & \textbf{Avg. Time} \\
\midrule
1 & \texttt{Codellama:70b}   & 6 & 21 & 31 & 4 & 0.60 & 0.22 & 0.60 & 0.32 & 0.40 & 97.37 \\
2 & \texttt{Llama2:70b}      & 10 & 52 & 0 & 0 & 0.16 & 0.16 & 1.00 & 0.28 & 1.00 & 104.05 \\
3 & \texttt{Llama3.3:70b}    & 7 & 1  & 51 & 3 & 0.94 & 0.88 & 0.70 & 0.78 & 0.02 & 82.63 \\
\bottomrule
\end{tabular}%
}
\end{table*}

To evaluate how very large language models perform in structural reasoning tasks, we benchmarked three representative 70-billion-parameter LLMs on the FlowConflict-ONOS dataset for conflict detection. As shown in Table~\ref{tab:conflict_ONOS_70b}, the results reveal a wide variance in effectiveness despite comparable scale. While Llama3.3:70b demonstrated strong overall performance—with 94\% accuracy, low false positive rate (0.02), and the highest F1-score (0.78) among the three—Codellama:70b and Llama2:70b underperformed significantly. Llama2:70b, despite perfect recall, produced 52 false positives, suggesting it over-predicted conflicts and failed to adhere to the prompt's strict field-level comparison criteria. Codellama:70b performed more moderately but still struggled with both recall and precision. Several factors contribute to the underperformance of large LLMs in this context. Large LLMs, trained on extensive and diverse datasets, tend to rely on semantic reasoning and pattern recognition. This predisposition can lead to overgeneralization, causing the models to infer conflicts where none exist, thereby increasing false positives. The training objectives of general-purpose large LLMs often do not align with the requirements of structured data interpretation. Without specific fine-tuning on tasks involving strict schema adherence and rule-based logic, these models struggle to accurately process and evaluate structured inputs like ONOS flow rules. However, the substantial computational resources required to run 70B parameter models can limit their practicality, especially when smaller models achieve comparable or superior performance with significantly lower resource consumption.
\vspace{-0.3cm}
\subsection{END-TO-END IBN REALIZATION USING NetIntent}
In this section, we show the implementation result of NetIntent. First we mention what LLM use choose for intent translation and conflict detection based on our LLM benchmark outcome. Then we report end-to-end delay of NetIntent for intent translation and intent activation on ODL and ONOS SDN controller.
\vspace{-0.3cm}
\subsubsection{SELECTED LLM FOR INTENT TRANSLATION AND CONFLICT DETECTION}
Based on the results of Table \ref{tab:Translation_ODL} and \ref{tab:conflict_ODL}, we choose QwQ:32b model for ODL for both intent translation and conflict detection tasks. As for ONOS, based on the results of Table \ref{tab:Translation_ODL} and \ref{tab:Conflict_ONOS}, we choose Codestral:22b model for intent translation and QwQ:32b model for conflict detection tasks. The context example was set to 3 for intent translation for both the controllers.
\vspace{-0.3cm}
\subsubsection{IMPLEMENTATION OUTCOME IN ODL AND ONOS SDN CONTROLLERS}
In Sec. VI-\ref{topology_setup}, we described the experimental topology used to evaluate our IBN framework. Using this setup, NetIntent accepts high-level natural language intents from the user. For simulation, we provided various intents such as \textit{``Deny packets originating from 10.0.0.1 destined for 10.0.0.4 using switch 1"} and \textit{``Do slicing at node 3 and ensure TCP packets destined for port 80, addressed to 10.0.0.3, are forwarded via port 2 using queue 0"} to test the end-to-end IBN workflow. NetIntent first applies Algorithm~\ref{alg:intent_translation} to translate the input intent into a structured JSON flow rule (based on target SDN controller, ODL or ONOS). This translated rule is then checked for potential conflicts against existing flow entries using Algorithm~\ref{alg:conflict_detection}. If a conflict is detected, NetIntent attempts to resolve it; if resolution is not possible, the intent is not activated, and the user is notified. If no conflict is found, the system proceeds to install the rule via the southbound API—RESTCONF for ODL or FlowRuleService for ONOS. Once installed, NetIntent verifies whether the flow is effective by inspecting its operational stats to confirm successful activation. The entire pipeline is fully automated, with the user interacting only at the intent level. Tables~\ref{tab:end_to_end_ODL} and~\ref{tab:end_to_end_ONOS} present the measured durations from the start of intent translation to confirmed rule activation for both ODL and ONOS deployments. These timings include translation, conflict detection, installation, and operational verification phases. While intent assurance was not included in the timing results (as it operates continuously in a closed-loop fashion), it was active in the background following Algorithm~\ref{alg:closed_loop_assurance} to ensure persistent verification of intent conformance.

\renewcommand{\arraystretch}{0.6}
\begin{table*}[t!]
\caption{End-to-end duration of NetIntent in ODL}
\label{tab:end_to_end_ODL}
\centering
\normalsize
\resizebox{\textwidth}{!}{%
\begin{tabular}{clccccccc}
\toprule
\textbf{Sl.} & \textbf{Intent} & \textbf{Intent Type} & \textbf{Existing Rule} & \textbf{Conflicting Rule} & \textbf{Translating LLM} & \textbf{Context} & \textbf{Detection LLM} & \textbf{E2E Time (s)} \\
\midrule
1 & In switch 4, install a firewall to block traffic from 10.0.0.2 to 10.0.0.4. & Security & 9 & 1 & \texttt{QwQ:32b} & 3 & \texttt{QwQ:32b} & 31.34 \\
1 & Drop all traffic from 10.0.0.9 on switch 2 while forwarding all other traffic normally. & Security & 2 & 1 & \texttt{QwQ:32b} & 3 & \texttt{QwQ:32b} & 29.75 \\
2 & Using openflow switch 1, forward UDP traffic on port 80 to 10.0.0.3 via interface 2, queue 0. & QoS & 8 & 1 & \texttt{QwQ:32b} & 3 & \texttt{QwQ:32b} & 24.02 \\
4 & If incoming traffic on interface 3 of node 3 is UDP to port 80, send via port 2, queue 1. & QoS & 4 & 0 & \texttt{QwQ:32b} & 3 & \texttt{QwQ:32b} & 25.47 \\
5 & If port 2 on switch 3 receives TCP to port 80, send via interface 3, queue 0. & QoS & 4 & 2 & \texttt{QwQ:32b} & 3 & \texttt{QwQ:32b} & 23.07 \\
6 & If port 2 on switch 3 receives UDP to port 80, pass via port 1, queue 0. & QoS & 4 & 2 & \texttt{QwQ:32b} & 3 & \texttt{QwQ:32b} & 35.10 \\
7 & In node 1, traffic to 10.0.0.2 should use port 3. & Forwarding & 8 & 1 & \texttt{QwQ:32b} & 3 & \texttt{QwQ:32b} & 33.70 \\
8 & In switch 2, traffic from port 1 should pass through port 3. & Forwarding & 2 & 1 & \texttt{QwQ:32b} & 3 & \texttt{QwQ:32b} & 20.10 \\
9 & Port 2 of switch 2 to 10.0.0.1 should use interface 4. & Forwarding & 2 & 0 & \texttt{QwQ:32b} & 3 & \texttt{QwQ:32b} & 26.50 \\
10 & In switch 4, traffic from 10.0.0.1 to 10.0.0.4 should use output interface 4. & Forwarding & 9 & 1 & \texttt{QwQ:32b} & 3 & \texttt{QwQ:32b} & 33.60 \\
\bottomrule
\end{tabular}%
}
\end{table*}

\renewcommand{\arraystretch}{0.6}
\begin{table*}[t!]
\caption{End-to-end duration of NetIntent in ONOS}
\label{tab:end_to_end_ONOS}
\centering
\normalsize
\resizebox{\textwidth}{!}{%
\begin{tabular}{clccccccc}
\toprule
\textbf{Sl.} & \textbf{Intent} & \textbf{Intent Type} & \textbf{Existing Rule} & \textbf{Conflicting Rule} & \textbf{Translating LLM} & \textbf{Context} & \textbf{Detection LLM} & \textbf{E2E Time (s)} \\
\midrule
1 & In switch 4, install a firewall to block traffic from 10.0.0.2 to 10.0.0.4. & Security & 9 & 1 & \texttt{Codestral:22b} & 3 & \texttt{QwQ:32b} & 30.34 \\
1 & Drop all traffic from 10.0.0.9 on switch 2 while forwarding all other traffic normally. & Security & 2 & 1 & \texttt{Codestral:22b} & 3 & \texttt{QwQ:32b} & 25.13 \\
2 & Using openflow switch 1, forward UDP traffic on port 80 to 10.0.0.3 via interface 2, queue 0. & QoS & 8 & 1 & \texttt{Codestral:22b} & 3 & \texttt{QwQ:32b} & 21.02 \\
4 & If interface 3 on node 3 receives UDP to port 80, pass via port 2, queue 1. & QoS & 6 & 0 & \texttt{Codestral:22b} & 3 & \texttt{QwQ:32b} & 28.56 \\
5 & If switch 3 receives TCP on port 2 to port 80, pass via interface 3, queue 0. & QoS & 6 & 2 & \texttt{Codestral:22b} & 3 & \texttt{QwQ:32b} & 22.07 \\
6 & If switch 3 receives UDP on port 2 to port 80, pass via port 1, queue 0. & QoS & 6 & 2 & \texttt{Codestral:22b} & 3 & \texttt{QwQ:32b} & 19.58 \\
7 & In node 1, traffic destined for 10.0.0.2 should use port 3. & Forwarding & 8 & 1 & \texttt{Codestral:22b} & 3 & \texttt{QwQ:32b} & 31.50 \\
8 & In switch 2, traffic from port 1 should pass through port 3. & Forwarding & 2 & 1 & \texttt{Codestral:22b} & 3 & \texttt{QwQ:32b} & 17.90 \\
9 & Traffic from port 2 of switch 2 to 10.0.0.1 should use interface 4. & Forwarding & 2 & 0 & \texttt{Codestral:22b} & 3 & \texttt{QwQ:32b} & 23.21 \\
10 & In switch 4, traffic from 10.0.0.1 to 10.0.0.4 should use output interface 4. & Forwarding & 9 & 1 & \texttt{Codestral:22b} & 3 & \texttt{QwQ:32b} & 29.20 \\
\bottomrule
\end{tabular}%
}
\end{table*}

\vspace{-0.3cm}
\section{LIMITATIONS AND FUTURE WORK}
\label{limitation_future_work}
Our benchmarks focus on representative natural language intents but do not encompass all possible configuration types or complex multi-intent scenarios. We also evaluated LLMs in their pretrained state, without task-specific fine-tuning, which may limit performance compared to customized models. Moreover, while NetIntent detects performance drift of intents, it does not address semantic drift~\cite{10979966} which is the presence of intents that are not specified by the user but are present in the network. Future research should explore fine-tuned models, as smaller, task-optimized models may perform well and be better suited for deployment in resource-constrained environments. Additionally, lightweight optimizations such as quantization, pruning, and knowledge distillation could further enhance the applicability of compact models. Beyond fine-tuning, customizing pretraining datasets and training strategies may improve model adaptability for IBN tasks and help reduce end-to-end latency. Investigating more efficient prompt designs could minimize translation errors and reduce false positives and negatives in conflict detection, thereby improving model reliability. Further efforts should examine LLMs’ reasoning capabilities to generate comprehensive explanations for decisions and output actions in IBN.


\vspace{-0.3cm}
\section{CONCLUSION}
\label{conclusion}
We benchmarked 33 open-source LLMs using our proposed IBNBench and found that their performance on isolated IBN tasks varies significantly. Crucially, task accuracy depends on prompt design, schema alignment, and in-context learning than on model size alone. Notably, models such as Gemma2-27B, QwQ-32B, QwQ-Fusion-32B, QwQ-Abliterated-32B, and Codestral-22B deliver consistent and high performance when guided by specialized prompting strategies. To enable LLM integration into a fully autonomous IBN pipeline, we introduced NetIntent, a unified, LLM-driven framework that automates the entire IBN lifecycle, including intent translation, policy activation, and assurance, across both ODL and ONOS SDN controllers. NetIntent coordinates LLM and non-LLM agents, enabling robust, natural language-driven intent realization with dynamic feedback and minimal human intervention. By releasing our open datasets, benchmarking results, and a practical NetIntent implementation, we establish a reproducible foundation for research on LLM-powered IBN. Our results demonstrate the feasibility of extensible, adaptive SDN automation, paving the way for future advances in intent-driven networking and LLM-native next-generation networking systems.

\vspace{-0.3cm}
\bibliographystyle{IEEEtran}
\bibliography{references}

\appendices
\section{EXAMPLE OF INTENT TRANSLATION}
\label{appendix:translated_intent}

\lstset{
  basicstyle=\scriptsize\ttfamily,
  breaklines=true,
  frame=none,
  columns=fullflexible,
  aboveskip=2pt,
  belowskip=2pt,
  lineskip=-3pt
}

\textbf{Intent:} Forward TCP traffic on port 80 destined for 10.0.0.3 via interface 2, assigning it to queue 0 to prioritize.

\begin{lstlisting}
ODL Flow Rule: {
  "flow-node-inventory:flow": [{
    "id": "1", "priority": 200, "table_id": 0,
    "flow-name": "tcp_port80_queue0_to_h3",
    "instructions": { "instruction": [{
      "order": 0, "apply-actions": {
        "action": [
          {"order": 0, "set-queue-action": {"queue-id": 0}},
          {"order": 1, "output-action": {"output-node-connector": "2"}}
        ]
      }
    }]},
    "match": {
      "ethernet-match": {"ethernet-type": {"type": 2048}},
      "ip-match": {"ip-protocol": 6},
      "ipv4-destination": "10.0.0.3/32",
      "tcp-destination-port": 80
    }
  }]
}
\end{lstlisting}

\begin{lstlisting}
ONOS Flow Rule: {
  "flows": [{
    "priority": 200, "timeout": 0, "isPermanent": "true",
    "deviceId": "of:0000000000000001",
    "treatment": { "instructions": [
      {"type": "QUEUE", "queueId": 0},
      {"type": "OUTPUT", "port": "2"}
    ]},
    "selector": { "criteria": [
      {"type": "ETH_TYPE", "ethType": "0x800"},
      {"type": "IP_PROTO", "protocol": 6},
      {"type": "IPV4_DST", "ip": "10.0.0.3/32"},
      {"type": "TCP_DST", "tcpPort": 80}
    ]}
  }]
}
\end{lstlisting}

\end{document}